\documentclass[reprint,notitlepage,groupedaddress,floatfix,superscriptaddress]{revtex4-2}
\usepackage[english]{babel}
\usepackage[utf8x]{inputenc}
\usepackage[T1]{fontenc}
\usepackage{graphicx}
\graphicspath{{Figures/}}
\usepackage{physics}
\usepackage[normalem]{ulem}
\usepackage{amsmath}
\usepackage{amsthm}
\usepackage{amssymb}
\usepackage{amsfonts}
\usepackage{mathtools}
\usepackage{enumerate}
\usepackage{color}
\usepackage[dvipsnames]{xcolor}
\usepackage{xifthen, xfrac}
\usepackage{cancel}
\usepackage[ddmmyyyy]{datetime}
\usepackage{euscript} 
\usepackage{mathbbol} 
\usepackage{epstopdf}
\usepackage{multirow,makecell,array}
\usepackage{float}
\usepackage{bbm}
\usepackage{dcolumn}
\usepackage{bm}
\usepackage{siunitx}
\usepackage[textwidth=1.2cm, textsize=scriptsize]{todonotes}

\begin{document}

\title{Semiclassical kinetic equations for composite bosons}

\author{A.~Kudlis}
\email{andrewkudlis@gmail.com}
\affiliation{Science Institute, University of Iceland, Dunhagi 3, IS-107, Reykjavik, Iceland}
\affiliation{Abrikosov Center for Theoretical Physics, MIPT, Dolgoprudnyi, Moscow Region 141701, Russia}
\author{I.~A.~Aleksandrov}
\email{i.aleksandrov@spbu.ru}
\affiliation{Department of Physics, Saint Petersburg State University, Universitetskaya Naberezhnaya 7/9, Saint Petersburg 199034, Russia}
\affiliation{Ioffe Institute, Politekhnicheskaya Street 26, Saint Petersburg 194021, Russia}
\author{Y. S. Krivosenko}
\affiliation{School of Physics and Engineering, ITMO University, Saint Petersburg 197101, Russia}
\author{I. A. Shelykh}
\affiliation{Science Institute, University of Iceland, Dunhagi 3, IS-107, Reykjavik, Iceland}

\begin{abstract}
We derive semiclassical Boltzmann equations describing thermalization of an ensemble of excitons due to exciton-phonon interactions taking into account the fact that excitons are not ideal bosons but composite particles consisting of electrons and holes. We demonstrate that with a standard definition of excitonic creation and annihilation operators, one faces a problem of the total particle number nonconservation and propose its possible solution based on the introduction of operators with angular momentum algebra. We then derive a set of kinetic equations describing the evolution of the excitonic density in the reciprocal space and analyze how the composite statistics of the excitons affects the thermalization processes in the system.
\end{abstract}

\allowdisplaybreaks

\maketitle

\newcommand{\numberthis}{\addtocounter{equation}{1}\tag{\theequation}}
\renewcommand{\vec}[1]{\vb*{#1}}
\newcommand{\vk}{\vec{k}}
\newcommand{\vvk}{\vec{\varkappa}}
\newcommand{\vkp}{\vec{\kappa}}
\newcommand{\vq}{\vec{q}}
\newcommand{\vp}{\vec{p}}

\renewcommand{\op}[1]{\hat{#1}}
\newcommand{\ham}{\op{H}}
\newcommand{\Ham}{\op{\mathcal{H}}}
\newcommand{\Aop}{\op{A}}
\newcommand{\Bop}{\op{B}}
\newcommand{\Cop}{\op{C}}
\newcommand{\Dop}{\op{D}}
\newcommand{\Xop}{\op{X}}
\newcommand{\Nop}{\op{N}}
\newcommand{\bop}{\op{b}}
\newcommand{\Lambdaop}{\op{\Lambda}}

\newcommand{\sumprime}{\mathop{{\sum}'\!}} 
\newcommand{\sumpprime}{\mathop{{\sum}''\!}} 
\newcommand{\prodprime}{\mathop{{\prod}'\!}} 

\newcommand{\vpt}{\tilde{\vp}}
\renewcommand{\abs}[1]{\left| {#1} \right|}
\newcommand{\parenth}[1]{\left( {#1} \right)}
\newcommand{\parenthl}[1]{\left( {#1} \right.}
\newcommand{\parenthr}[1]{\left. {#1} \right)}
\newcommand{\brackets}[1]{\left[ {#1} \right]}
\newcommand{\dkk}{\sigma} 
\newcommand{\Dkk}{\Delta} 
\newcommand{\ac}[1]{\{ {#1} \} } 
\newcommand{\df}{\delta'}
\newcommand{\Df}{\Delta'}
\newcommand{\culine}[2]{\textcolor{#2}{\uline{\textcolor{Black}{#1}}}}
\newcommand{\cuwave}[2]{\textcolor{#2}{\uwave{\textcolor{Black}{#1}}}}
\colorlet{palegray}{CadetBlue!50!Black}
\newcommand{\mean}[1]{\langle {#1} \rangle}
\newcommand{\eqtext}[1]{\overset{#1}{=\joinrel=}}
\newcommand{\eqtextl}[2]{ 
  \overset{#2}{
  \foreach \y in {1,...,{#1}}{
    =\joinrel
    }
  =
  }
}
\newcommand{\eqdef}{\eqtext{\mathrm{def}}}
\newcommand{\ab}{a_{\text{b}}}
\newcommand{\Paulikkkk}{
\lambda
\begin{pmatrix}
  \vk & \vk \\
  \vk & \vk
\end{pmatrix}
}
\newcommand{\diff}{\mathrm{d}}
\newcommand{\elec}{e}
\newcommand{\hole}{h}
\newcommand{\e}{\mathrm{e}}
\newcommand{\chircp}{\widetilde{\chi}_{1s}}
\newcommand{\hc}{\text{h.c.}}
\newcommand{\add}{\mathrel{{+}{=}}}
\newcommand{\subtract}{\mathrel{{-}{=}}}
\newcommand{\alphain}{\alpha^{\downarrow}}
\newcommand{\alphaout}{\alpha^{\uparrow}}
\newcommand{\wt}[1]{\widetilde{#1}}
\newlength{\hstep}
\setlength{\hstep}{1.5em}
\newcommand{\hence}{\ \Rightarrow \ }
\newcommand{\hencer}{\ \Rightarrow }
\newcommand{\hencel}{\Rightarrow \ }
\newcommand{\nph}{\hat{n}_{\vq}}
\newcommand{\fvk}{f_{\vk}}
\newcommand{\fdvk}{f^{\dagger}_{\vk}}
\newcommand{\fvkq}{f_{\vk+\vq}}
\newcommand{\fdvkq}{f^{\dagger}_{\vk+\vq}}
\newcommand{\Bpf}{B^{+}_{\text{f}}}
\newcommand{\Hhop}{\hat{H}_{\text{hop}}}
\newcommand{\ekkkk}{e^{\dagger}_{\vk_1} e_{\vk_2} e^{\dagger}_{\vk_3} e_{\vk_4}}
\newcommand{\ccnice}{c^\dagger_{\vk+\frac{\vvk}{2}} c^{\vphantom{\dagger}}_{\vk+\frac{\vvk}{2}}}
\newcommand{\vvnice}{v^\dagger_{\vk-\frac{\vvk}{2}} v^{\vphantom{\dagger}}_{\vk-\frac{\vvk}{2}}}
\newcommand{\cvnice}{c^\dagger_{\vk+\frac{\vvk}{2}} v^{\vphantom{\dagger}}_{\vk-\frac{\vvk}{2}}}
\newcommand{\vcnice}{v^{\dagger}_{\vk-\frac{\vvk}{2}} c^{\vphantom{\dagger}}_{\vk+\frac{\vvk}{2}}}
\newcommand{\sps}{\mathfrak{S}}
\newcommand{\sumk}[1][]
{
\ifthenelse{\isempty{#1}}{\sum}{\sum\limits_{#1}}
}
\newcommand{\mcS}{\op{\mathcal{S}}}
\newcommand{\newS}{\op{\mathbb{S}}}
\newcommand{\news}{\mathbb{s}}
\newcommand{\newxsign}{\mathbb{x}}
\newcommand{\newx}{\op{\newxsign}}
\newcommand{\hatphi}{\hat{\Phi}}
\newcommand{\Hnewhop}{\mathbb{H}_{\text{hop}}}
\newcommand{\Hnewp}{\mathbb{H}^+}
\newcommand{\Hnewm}{\mathbb{H}^-}
\newcommand{\Dnew}{\mathcal{D}}
\newcommand{\area}{\mathcal{A}}
\newcommand{\upsign}{\uparrow}
\newcommand{\downsign}{\downarrow}
\newcommand{\Bdir}{\EuScript{B}^{\upsign}}
\newcommand{\Binv}{\EuScript{B}^{\downsign}}
\newcommand{\pumpp}{\mathcal{P}}

\section{Introduction}
Wannier--Mott excitons are composite quasiparticles formed by negatively charged conductance-band electrons and positively charged valence-band holes bound together by the Coulomb attraction~\cite{HaugKoch2009, Knox1963}.
Excitons possess an integer spin and thus in the diluted limit demonstrate clear signatures of the bosonic behavior~\cite{Kavokin2017_OxfPr}, revealing such phenomena as excitonic stimulated scattering~\cite{Tassone1999,Huang2000,Butov2001} and even excitonic Bose-Einstein condensation~\cite{Butov2002,Morita2022}. 

The temporal dynamics of excitonic ensembles can be experimentally studied by using time-resolved photoluminescence~\cite{Gurioli1998_PRB, Vinattieri1994_PRB, Eccleston1991_PRB, Hoyer2005_PRB, Cadiz2018_APL}. A standard theoretical method for modeling the experimental data implies solving a set of coupled semiclassical Boltzmann equations for excitonic occupancies in the reciprocal (momentum) space~\cite{Snoke2011_AnnPhys, Savenko2011_PRB, Snoke1991_PRB, Snoke1992_PRB, OHara2000_PRB, Ivanov1997_PRE, Potma1998_JChemPhys, Selig2018_2DMat, Chen2022_PhysRevRes, Piermarocchi1997_PRB, Zhang1997_JPCM, Piermarocchi1996_PRB, Selbmann1996_PRB, Golub1996_JPCM, Basu1992_PRB, Lee1986_PRB}. In the most general case, this set should account for a variety of mechanisms governing the redistribution of the excitons in the reciprocal space, which include impurity scattering, exciton-exciton collisions, and interaction with phonons. The latter mechanism is of specific importance as it leads to the thermalization of the excitonic system. 

Kinetic equations used for description of the excitonic dynamics routinely use the representation of an exciton as an elementary boson lacking any internal structure. However, being composed of individual fermions, excitons can be treated as bosons only if the average distance between them is much larger than the exciton Bohr radius $a_\text{B}$, so that the following condition is satisfied:
\begin{equation}
n_\text{X} a_\text{B}^d \ll 1,
\label{eq:condition}
\end{equation}
where $n_\text{X}$ is the concentration of the excitons and $d=1,2,3$ is the spatial dimension of the system.

The composite nature of excitons leading to deviations from a truly bosonic behavior has attracted attention since the early 2000s.
An efficient approach based on the concept of composite bosons (\textit{cobosons}) was developed in Refs.~\cite{Combescot2002_EurohysLett, Combescot2008_PhysRep, Combescot2007_PRB} by Combescot and co-authors, and the effects of the Pauli exclusion principle at the level of individual electrons and holes were further analyzed by Thilagam~\cite{Thilagam2015_APA} and Katzer and co-authors~\cite{Katzer2023_arxiv}. 

However, the problem of a derivation of semiclassical kinetic equations describing the time-resolved dynamics of the thermalization process in the excitonic system which accounts for the composite statistics of the excitons has been never been analyzed, to the best of our knowledge. Here we present an attempt to construct the corresponding theory. Our main goal is to obtain a general version of the Boltzmann kinetic equations incorporating the statistical effects regarding the exciton composite structure. To this end, we further advance the approach based on the use of the master equation for the excitonic density matrix, presented in Refs.~\cite{Savenko2011_PRB,Golub1996_JPCM,Piermarocchi1996_PRB,Combescot2008_PhysRep}, generalizing it for the case of a nonbosonic statistics of the particles.

The paper is structured as follows. In Sec.~\ref{sec:Theory}, we recall the standard derivation of semiclassical Boltzmann equations (BBEqs) for exciton-phonon interaction, discuss the related problem of the conservation of the total number of the composite excitons, introduce the excitonic operators in terms of high angular momentum algebra, and derive the corresponding semiclassical Boltzmann equations (BEqs) conserving the total number of the particles. In Sec.~\ref{sec:RnD}, we perform numerical simulations of the system evolution for both a simple model case clearly revealing the role of the effects of the composite statistics and a realistic geometry corresponding to excitonic relaxation in 1D GaAs quantum wire. The concluding remarks are then provided in Sec.~\ref{sec:Conclusion}.

\section{Theoretical model \label{sec:Theory}}

\subsection{Model Hamiltonian \label{sec:Theory_BEqs}}

A system of excitons interacting with acoustic phonons can be described by the following model Hamiltonian:
\begin{equation}
\hat{H}=\hat{H}_0+\hat{H}_{\rm SR}.
\label{eq:Ham1}
\end{equation}
Here the first term corresponds to free excitons and phonons, 
\begin{equation}
\hat{H}_0=\sum\limits_{\vk} \varepsilon_{\vk} \, \hat{N}_{\vk} + \sum\limits_{\vk} \varepsilon_{\vk}^{\rm  ph} \, \hat{b}^{\dagger}_{\vk} \hat{b}_{\vk} \label{eq:Ham0}
\end{equation}
with $\hat{N}_{\vk}$ being the operators of the number of excitons in a state with momentun $\bm{k}$ and energy $\varepsilon_{\vk}$, while  $\hat{b}^{\dagger}_{\vk}$ and $\hat{b}_{\vk}$ the bosonic operators of creation and annihilation of phonons with momentun $\bm{k}$ and energy $\varepsilon_{\vk}^{\rm  ph}$, and the term 
\begin{equation}
\hat{H}_{\rm SR}=\sum\limits_{\vk,\vq} D(\vq) \left( \Xop^\dagger_{\vk+\vq} \Xop_{\vk} \bop_{\vq}+ \Xop^\dagger_{\vk} \Xop_{\vk+\vq} \bop^{\dagger}_{\vq}\right) \label{eq:HamSR}
\end{equation}
is the Hamiltonian describing the interaction between the excitonic system and the phonon reservoir leading to emission and absorbtion of the phonons and corresponding redistrubution of excitons in the reciprocal space. 

In Eq.~\eqref{eq:HamSR} $\Xop^\dagger_{\vk}$ and $ \Xop_{\vk}$ stand for the excitonic creation and annihilation operators, respectively. In the diluted limit, when the condition~\eqref{eq:condition} is satisfied, the excitons behave as bosons, so that $\big [\Xop^\dagger_{\vq}, \Xop_{\vk}\big]=\delta_{\vk\vq}$ leading to
\begin{equation}
\hat{N}_{\vk}=\Xop^\dagger_{\vk}\Xop_{\vk}, \label{eqN}   
\end{equation}
and the dynamics of the excitonic occupancies $n_{\bm{k}}^{\rm X}$ defined as the mean values of the corresponding operators~\eqref{eqN} in Born-Markov approximation~\cite{Carmichael2007,Kavokin2003,Savenko2011_PRB} is described by a set of semiclassical Boltzmann equations which involve the terms corresponding to the final state bosonic stimulation:
\begin{widetext}
\begin{multline}
\partial_t n_{\bm{k}}^{\rm X} =\!\!\!\!\!\!\!\! \sum_{\bm{q}, \varepsilon_{\bm{k}} < \varepsilon_{\bm{k} + \bm{q}}}\!\!\!\!\!\!\!\!
W_{\bm{k},\bm{k}+\bm{q}} \!\!\left[n_{\bm{k} + \bm{q}}^{\rm X}(n^{\rm ph}_{\bm{q}} + 1)(n_{\bm{k}}^{\rm X} + 1) 
- n_{\bm{k}}^{\rm X} n^{\rm ph}_{\bm{q}} (n_{\bm{k} + \bm{q}}^{\rm X} + 1)\right]\\
+ \!\!\!\!\!\!\!\! \sum_{\bm{q}, \varepsilon_{\bm{k}} > \varepsilon_{\bm{k} + \bm{q}}}\!\!\!\!\!\!\!\! 
W_{\bm{k}+\bm{q},\bm{k}} \left[ n_{\bm{k} + \bm{q}}^{\rm X} n^{\rm ph}_{-\bm{q}} (n_{\bm{k}}^{\rm X} + 1)
- n_{\bm{k}}^{\rm X} (n^{\rm ph}_{-\bm{q}} + 1)(n_{\bm{k} + \bm{q}}^{\rm X} + 1)\right].
\label{eq:BBeqs_tot}
\end{multline}
\end{widetext}
Here $n_{\vq}^{\rm ph}$ are the occupation numbers of the acoustic phonons with momentum $\vq$ defined by a standard Bose-Einstein distribution function and $W_{\bm{k},\bm{k}+\bm{q}}$ are the transition rates determined by the parameters of the Hamiltonian and line broadenings.

However, as we show below, taking into account the composite nature of the excitons makes the relation~\eqref{eqN} obsolete as its use leads to the problem of the particle nonconservation in the process of exciton-phonon scattering.

\subsection{The problem of total number conservation for composite particles \label{sec:MBEqs}}

The system of equations~\eqref{eq:BBeqs_tot} ensures the conservation of the total number of the excitons, which represents a direct consequence of the commutativity between the system Hamiltonian and the operator of the total number of the excitons defined as 
\begin{equation}
\Nop = \sum\limits_{\vk} \Xop^{\dagger}_{\vk} \Xop_{\vk} = \sum\limits_{\vk} \Nop_{\vk}, \label{Ntot}
\end{equation}
for the case when the operators $\Xop^{\dagger}_{\vk}$ and $\Xop_{\vk}$ obey a standard bosonic algebra. 

This is not the case, however, if the composite nature of the excitons consisting of individual electrons and holes is accounted for~\cite{Combescot2008_PhysRep}.

To show this, let us consider the standard expression for exciton creation operator defined as~\cite{Combescot2008_PhysRep}
\begin{equation}
\Xop^{\dagger}_{\vk} = \sum\limits_{\vvk} \widetilde{\chi} (\vvk) \, \hat{\elec}^{\dagger}_{{\vk}/{2}+\vvk} \, \hat{\hole}^{\dagger}_{{\vk}/{2}-\vvk},
\label{eq:X-operator-def}
\end{equation}
where $\hat{\elec}^{\dagger}_{\vk}$ and $\hat{\hole}^{\dagger}_{\vk}$ denote the electron and hole creation operators that obey {\it fermionic} anticommutation relations. In Eq.~\eqref{eq:X-operator-def}, $\vvk$ and $\vk$ denote the electron-hole relative and center-of-mass momenta, respectively, and $\widetilde{\chi} (\vvk)$ is the exciton relative-motion wave function in the momentum space. The latter is assumed to correspond to the ground state ($1s$) of the bound electron-hole system as we neglect here possible excitations of the internal degrees of freedom.

The commutator involving the cobosonic operators $\Xop_{\vk}$ and $\Xop^{\dagger}_{\vk}$ defined in Eq.~\eqref{eq:X-operator-def} is known to differ from the Kronecker delta $\delta_{\vk\vk'}$, which is valid for bosons, and reads~\cite{Combescot2008_PhysRep}
\begin{equation}
\comm\big{\Xop_{\vk}}{\Xop^\dagger_{\vk'}} =
\delta_{\vk\vk'} - \hat{\Dkk}_{\vk,\vk'},
\label{eq:Dkk_def}
\end{equation}
where 
\begin{widetext}
\begin{equation}
\hat{\Dkk}_{\vk,\vk'} = \sum\limits_{\vvk} \Big [ \widetilde{\chi} (\vvk - \vk/2) \widetilde{\chi}^* (\vvk - \vk'/2) \, \hat{\elec}^{\dagger}_{\vvk - (\vk-\vk')/2} \, \hat{\elec}_{\vvk + (\vk-\vk')/2} + \widetilde{\chi} (\vvk + \vk/2) \widetilde{\chi}^* (\vvk + \vk'/2) \, \hat{\hole}^{\dagger}_{-\vvk - (\vk-\vk')/2} \, \hat{\hole}_{-\vvk + (\vk-\vk')/2} \Big].
\label{eq:deltakk}
\end{equation}
\end{widetext}
By a direct analogy with the case of bosonized excitons, one can introduce the operator of the total exciton number by Eq.~\eqref{Ntot}. However, the explicit calculation shows that the deviation from the bosonic statistics described by the term $\hat{\Dkk}_{\vk, \vk'}$ yields
\begin{multline}
\!\!    [\hat{H},\hat{N}]=\sum\limits_{\vk,\bm{p}} \left(\varepsilon_{\vk}-\varepsilon_{\bm{p}}\right)\Xop^{\dagger}_{\bm{p}}\hat{\Dkk}^{}_{\bm{p},\bm{k}}\Xop^{}_{\bm{k}}
    +\sum\limits_{\vk,\bm{p},\bm{q}}D(\bm{q}) \\ \times\left(b_{\bm{q}}+b_{-\bm{q}}^{\dagger}\right)\left(\Xop^{\dagger}_{\bm{p}}\hat{\Dkk}^{}_{\bm{p},\bm{k}+\bm{q}}\Xop^{}_{\bm{k}}-\Xop^{\dagger}_{\bm{p}}\hat{\Dkk}^{}_{\bm{p}-\bm{q},\bm{k}}\Xop^{}_{\bm{k}}\right).\!\!\!\!
\end{multline}
From the physical viewpoint, this means that the total number of the particles in the system is not anymore a conserving quantity. This is a severe problem requiring a proper redefinition of the excitonic operators.

\subsection{Definition of the excitonic operators}

To simplify the following analysis, let us start with a two-level model describing the phonon-assisted redistribution of particles between a pair of states and governed by the following model Hamiltonian:
\begin{multline}
    \!\!\!\hat{H}=\varepsilon_1 \hat{N}_1+\varepsilon_2 \hat{N}_2 + \varepsilon^{\rm{ph}}\hat{b}^{\dagger}\hat{b}\\+D(\hat{\tilde{X}}^\dagger_{2}\hat{\tilde{X}}^{}_{1}\hat{b}+\hat{\tilde{X}}^\dagger_{1}\hat{\tilde{X}}^{}_{2}\hat{b}^{\dagger}),
\end{multline}
where $\hat{N}_1$ and $\hat{N}_2$ are the particle number operators of the two states, and $\hat{\tilde{X}}^{\dagger}_{1}$ and $\hat{\tilde{X}}^{\dagger}_{2}$ are corresponding excitonic creation operators. We assume that the energy levels are not degenerate and obey $\varepsilon_2=\varepsilon_1+\varepsilon^{\rm{ph}}>\varepsilon_1$. 

The condition $[\hat{H},\hat{N}]=0$ is satisfied if the operators have the following algebraic properties:

a) $[\hat{X}_{i},\hat{X}_{j}]=0$, $[\hat{X}_{i},\hat{X}^\dagger_{j}]=\delta_{ij}$, and $\hat{N}_i=\hat{X}^\dagger_{i} \hat{X}_i$. This corresponds to the already considered bosonic case.

b) $\{\hat{X}_{i},\hat{X}_{j}\}=0$, $\{\hat{X}_{i},\hat{X}^\dagger_{j}\}=\delta_{ij}$, and $\hat{N}_i=\hat{X}^\dagger_{i} \hat{X}_i$, where $\{...\}$ stands for an anticommutator. This corresponds to the case of fermionic particles.

c) $[\hat{X}_{i},\hat{X}_{j}]=0$, $[\hat{X}_{i},\hat{X}^\dagger_{j}]=\delta_{ij}(1-\hat{N}_i/J_i)$, $[\hat{X}_i,\hat{N}_i]=\hat{X}_i/\sqrt{2J_i}$, and $[\hat{X}^\dagger_i,\hat{N}_i]=-\hat{X}^\dagger_i/\sqrt{2J_i}$. In this case, the excitonic operators can be defined via the operators of angular momentum $\hat{\boldsymbol{J}}$:
\begin{align}
    \hat{X}^\dagger_{i}=\dfrac{\hat{J}_i^{+}}{\sqrt{2J_i}}, \quad    \hat{X}^{}_{i}=\dfrac{\hat{J}_i^{-}}{\sqrt{2J_i}},\label{eqn:new_operators}
\end{align}
where $J_i$ are positive integer or half-integer numbers and 
\begin{equation}
\hat{N}_i=\hat{J}_{i}^z+J_i.
\end{equation}
In the case of nonbosonic excitons, the physical meaning of the parameter $J_i$ can be established as following. We note that $J_i$ is the absolute value of the angular momentum, which has $2J_i+1$ different projections. One of these projections corresponds to the vacuum state, so the maximal possible number of excitons in a given state $i$ is nothing but 
\begin{equation}
N_{i,{\rm max}}=2J_i.
\end{equation}
The Hamiltonian rewritten in terms of the angular momentum operators reads
\begin{multline}
    \hat{H}=\varepsilon_1\left(\hat{J}^z_1+\dfrac{N_{1,\rm max}}{2}\right)+\varepsilon_2\left(\hat{J}^z_2+\dfrac{N_{2,\rm max}}{2}\right)\\ + \varepsilon^{\rm{ph}}\hat{b}^{\dagger}\hat{b}+\tilde{D}(J_2^{+}J_1^{-}\hat{b}+J_1^{+}J_2^{-}\hat{b}^{\dagger}),\label{eq:H_2level}
\end{multline}
where $\tilde{D}=D/(2\sqrt{J_1 J_2})$. Direct check shows that it indeed commutes with $\hat{J}^z=\hat{J}_1^z + \hat{J}_2^z$, which ensures the conservation of the total number of the excitons $\hat{N}=\hat{N}_1+\hat{N}_2 = \hat{J}_1^z+\hat{J}_2^z+J_1+J_2$.

We can relate the maximal possible occupancy of a single level to the microscopic parameter characterizing an exciton, i.e., its Bohr radius. In our further discussion, we will make an assumption that the change of momentum $\vk$ does not affect the exciton wave function, so that in terms of the maximal occupancy, all states in our system are equivalent to the state~$\vk$. 

Let us consider $\vk = 0$ and estimate now the maximum occupation number $N_{\rm max}$. To this end, we inspect the connection between the commutation relation~\eqref{eq:Dkk_def} for the excitonic operators written in terms of the standard coboson operators for $\bm{k}=0$ [see Eq.~\eqref{eq:X-operator-def}] and its counterpart for the excitonic operators expressed via the angular momentum algebra,
\begin{equation}
    [\hat{X}_0^\dagger,\hat{X}_0]=1-\dfrac{2\hat{N}_0}{N_{\rm max}},
\end{equation}
one immediately gets:
\begin{align}
\dfrac{2 \hat{N}_0}{N_{\rm max}}=\hat{\Delta}_{0}=\sum_{\vvk} |\widetilde{\chi} (\vvk)|^2( \hat{n}^{e}_{\vvk} + \hat{n}^{h}_{-\vvk}).\label{eq:comb_eq}
\end{align}
We can now construct a state consisting $N_0$ excitons as:
\begin{equation}
\ket{N_0}=(\hat{X}_0^{\dagger})^{N_0}\ket{\varnothing}   
\end{equation}
and average Eq.\eqref{eq:comb_eq} with to get
\begin{align}
 &\dfrac{2 N_0}{N_{\rm max}}= \dfrac{\bra{N_0}\hat{\Delta}_0\ket{N_0}}{\braket{N_0}{N_0}}=\label{Average}\\ \nonumber&=\dfrac{\bra{\varnothing} (\hat{X}_0)^{N_0}[\hat{\Delta}_0,(\hat{X}_0^{\dagger})^{N_0}]\ket{\varnothing}}{\bra{\varnothing} (\hat{X}_0)^{N_0}(\hat{X}_0^{\dagger})^{N_0}]\ket{\varnothing}}=N_0C, 
\end{align}
where $C$ is the following overlap integral
\begin{multline}
   C=\int \! {\rm d}\mathbf{r}_{\alpha_1} {\rm d}\mathbf{r}_{\alpha_2} {\rm d}\mathbf{r}_{\beta_1} {\rm d}\mathbf{r}_{\beta_2} \\
\times\phi^*_0(\mathbf{r}_{\alpha_2}, \mathbf{r}_{\beta_2})\phi^*_0(\mathbf{r}_{\alpha_1}, \mathbf{r}_{\beta_1})
\phi_0(\mathbf{r}_{\alpha_1}, \mathbf{r}_{\beta_1}) \phi_0(\mathbf{r}_{\alpha_2}, \mathbf{r}_{\beta_2}),\label{eq:second_av_rel}
\end{multline}
and for the commutator in the numerator of Eq.\eqref{Average} we have used the following relation \cite{COMBESCOT2008215}:
\begin{equation}
    [\hat{\Delta}_0,(\hat{X}_0^{\dagger})^{N_0}]\approx N_0(\hat{X}^{\dagger}_0)^{N_0}C,
\end{equation}
which is valid in the linear order by $N_0$. Here, $\phi_0(\mathbf{r}_{\alpha 1}, \mathbf{r}_{\beta 1})$ is the corresponding exciton wave function in the real-space representation. 

The integral \eqref{eq:second_av_rel} can be calculated analytically, if one represents the excitonic wavefunction as a product of the center-of-mass and relative-motion components,
\begin{align}
\phi_0(\mathbf{r}_\alpha, \mathbf{r}_\beta) = \frac{e^{{\rm i} \mathbf{K}_0 \mathbf{R}_{\alpha\beta}}}{\sqrt{S}} \varphi_0(\mathbf{r}_{\alpha\beta}),
\end{align}
where $\mathbf{R}_{\alpha\beta}$ and $\mathbf{r}_{\alpha\beta}$ are the center-of-mass and relative coordinates of the $e$-$h$ pair and $\mathbf{K}_0$ is the total momentum of the coboson, $S$ is an area of the sample, and then uses a hydrogen-type approximation for the wavefunction of the relative motion $\varphi_0(\mathbf{r}_{\alpha\beta})$:  
\begin{equation}
\varphi_0(\mathbf{r}_{\alpha\beta})=\sqrt{\frac{8}{\pi a_{\rm B}^2}}e^{-2r/a_{\rm B}},
\end{equation}
where $a_{\rm B}$ is the corresponding Bohr radius. 

In this case, for the maxamal possible excitonic occupancy we obtain
\begin{align}
    N_{\rm max}=\dfrac{5}{4\pi}\dfrac{S}{a_{\rm B}^2}.
    \label{eq:N0max}
\end{align}
Note that accorging to this expression the condition~\eqref{eq:condition} is equivalent to $N_0 \ll N_{\rm max}$, which exactly corresponds to the limit when bosonization becomes possible.

In what follows, we will use this final estimate in our numerical simulations. 

\subsection{Equations of motion}

Let us now derive kinetic equations describing the dynamics of the level occupancies. We start with a simple two-level case and then generalize our derivation for a more realistic case involving excitons distributed in $\vk$ space. 

We will work in the interaction picture, where the corresponding coupling Hamiltonian reads
\begin{multline}
    \hat{H}_{\rm SR}=\tilde{D}e^{-{\rm i}\hat{H}_{0}t/\hbar}(\hat{J}_2^{+}\hat{J}_1^{-}\hat{b}+\hat{J}_1^{+}\hat{J}_2^{-}\hat{b}^{\dagger})e^{{\rm i}\hat{H}_{0}t/\hbar}\\
    =\tilde{D}\big [\hat{J}_2^{+}\hat{J}_1^{-}\hat{b}e^{{\rm i}(\varepsilon_2-\varepsilon_1-\varepsilon^{\rm{ph}})t/\hbar}+\hat{J}_1^{+}\hat{J}_2^{-}\hat{b}^{\dagger}e^{-{\rm i}(\varepsilon_2-\varepsilon_1-\varepsilon^{\rm{ph}})t/\hbar}\big]\\
    = \tilde{D} (\hat{J}_2^{+}\hat{J}_1^{-}\hat{b}+\hat{J}_1^{+}\hat{J}_2^{-}\hat{b}^{\dagger}) \equiv \hat{H}^{-}+\hat{H}^{+},
\end{multline}
where
\begin{equation}
\hat{H}_0=\varepsilon_1\left(\hat{J}^z_1+\dfrac{N_{1,\rm max}}{2}\right)+\varepsilon_2\left(\hat{J}^z_2+\dfrac{N_{2,\rm max}}{2}\right).
\end{equation}
The Liouville–von Neumann equation for the density matrix in the interaction picture then has the form
\begin{equation}
    {\rm i} \partial_t\hat{\chi}=\dfrac{1}{\hbar}\big[\hat{H}_{\rm SR},\hat{\chi}\big],\label{eqn:den_mtr}
\end{equation}
which after formal time integration yields
\begin{align}\label{eqn:sol_dens_matr_1}
    \hat{\chi}(t)=\hat{\chi}(-\infty)+\dfrac{1}{{\rm i}\hbar}\int\limits_{-\infty}^t \! {\rm d}t'\big[\hat{H}_{\rm SR}(t'),\hat{\chi}(t')\big].
\end{align}

Assuming that the interaction is adiabatically switched, $[\hat{{H}}_{\rm SR},\hat{{\chi}}(-\infty)]=0$, we can substitute Eq.~\eqref{eqn:sol_dens_matr_1} into Eq.~\eqref{eqn:den_mtr} and thus obtain
\begin{equation}\label{eq:chi_comm}
    \partial_t\hat{\chi} (t)=-\dfrac{1}{\hbar^2}\int\limits_{-\infty}^t \! {\rm d}t'\big[\hat{H}_{\rm SR}(t),\big[\hat{H}_{\rm SR}(t'),\hat{\chi}(t')\big]\big].
\end{equation}
Now we employ the Born-Markov approximation by replacing
$\hat{\chi}(t')\rightarrow\hat{\chi}(t)$, which allows one to explicitly integrate over $t'$ and find the following energy-conservation relation:
\begin{multline}
\dfrac{1}{\hbar} \int\limits_{-\infty}^{t} \! {\rm d}t' \hat{H}_{\rm SR}(t') = \tilde{D}(\hat{J}_2^{+}\hat{J}_1^{-}\hat{b}+\hat{J}_1^{+}\hat{J}_2^{-}\hat{b}^{\dagger})\delta(\varepsilon_2 - \varepsilon_1 - \varepsilon^{\rm  ph})\\
\approx\dfrac{\tilde{D}}{\Delta_E}(\hat{J}_2^{+}\hat{J}_1^{-}\hat{b}+\hat{J}_1^{+}\hat{J}_2^{-}\hat{b}^{\dagger})=\frac{1}{\Delta_E}\left(\hat{H}^-+ \hat{H}^+\right).
\end{multline}
In the last line, we replaced the Dirac delta function with an inverse broadening of the levels according to the conventional regularization procedure~\cite{Carmichael2007,Kavokin2003}. 

Neglecting the anomalous terms such as those containing $\hat{H}^+\hat{h}^+$, we finally obtain the following master equation for the density matrix:
\begin{equation}\label{eqn:dens_matr_final_eq}
   \partial_t\hat{\chi}=\dfrac{1}{\hbar\Delta_E}\Big(\comm\big{\hat{H}^+}{\comm\big{\hat{\chi}}{\hat{H}^-}}+\comm\big{\hat{H}^-}{\comm\big{\hat{\chi}}{\hat{H}^+}}\Big).
\end{equation}

To derive a kinetic equation for any physical observable which does not explicitly depend on time, it suffices just to take a trace of the master equation~\eqref{eqn:dens_matr_final_eq} with the corresponding operator. For the operators $\hat{J}^z_i$ determining the occupancies of the energy levels, one finds
\begin{multline}\label{eq:motion_of_Jz}
    \partial_t  \big \langle \hat{J}^z_1 \big\rangle = - \partial_t  \big \langle \hat{J}^z_2\big\rangle=\frac{2}{\hbar\Delta_E} \, \text{Re} \, \big\langle \comm\big{\hat{H}^-}{\comm\big{\hat{J}_1^z}{\hat{H}^+}} \big\rangle\\
   =\frac{2\tilde{D}^2}{\hbar\Delta_E} \left[\langle\hat{J}_2^+\hat{J}_2^-\rangle\langle\hat{J}_1^-\hat{J}_1^+\rangle\left(1+n^{\rm ph}\right)\right.\\
   +\left.\langle\hat{J}_1^+\hat{J}_1^-\rangle\langle\hat{J}_2^-\hat{J}_2^+\rangle n^{\rm ph}\right], 
\end{multline}
where the angular brackets denote the mean value $\langle \hat{J}^z_1 \rangle = \mathrm{Tr} \, (\hat{J}^z_1 \hat{\chi})$. We used Born approximation, i.e., represented the density operator as a direct product of the density operators corresponding to the excitonic states and phonons, $\hat{\chi}=\hat{\chi}_\text{X}\otimes\hat{\chi}_{\rm ph}$ and introduced $n^{\rm ph}=\langle b^\dagger b\rangle$ being the phonon occupancy defined by a standard Bose distribution function. Note that if one uses bosonic operators instead of $\hat{J}^\pm$, one obtains from \eqref{eq:motion_of_Jz} the standard bosonic Boltzmann equations.

Using the angular momentum algebra
\begin{align}
\hat{J}_i^{+}\hat{J}_i^{-}&=\hat{\boldsymbol{J}}_i^2-(\hat{J}_i^z)^2+\hat{J}_i^z\nonumber \\
&=\hat{\boldsymbol{J}}_i^2-(\hat{N}_i-J_i)(\hat{N}_i-J_i-1), \\
\hat{J}_i^{-}\hat{J}_i^{+}&=\hat{\boldsymbol{J}}_i^2-(\hat{J}_i^z)^2-\hat{J}_i^z \nonumber \\
&=\hat{\boldsymbol{J}}_i^2-(\hat{N}_i-J_i)(\hat{N}_i-J_i+1),
\end{align}
we obtain 
\begin{align}
    &\langle \hat{J}_i^{+}\hat{J}_i^{-}\rangle=N_i(N_{i,{\rm max}}+1-N_i),\label{eq:approx_laderpm}\\
    &\langle \hat{J}_i^{-}\hat{J}_i^{+}\rangle=(N_i+1)(N_{i,{\rm max}} - N_i),\label{eq:approx_ladermp}
\end{align}
where we used an approximation $\langle\hat{N}_i^2\rangle\approx\langle\hat{N}_i\rangle^2=N_i^2$, which is valid if the statistics of the occupancies is close to Poissonian. This allows one to derive a closed-form system of equations governing the occupation numbers:
\begin{widetext}
\begin{align}\label{eqn:final_two_levels}
    &\partial_t N_1=W_{12}\left\{N_2\left(1-\dfrac{N_2-1}{N_{2,{\rm max}}}\right)\left[1+N_1\left(1-\dfrac{N_1+1}{N_{1,{\rm max}}}\right)\right]\left(1+n^{\rm ph}\right)-N_1\left(1-\dfrac{N_1-1}{N_{1,{\rm max}}}\right)\left[1+N_2\left(1-\dfrac{N_2+1}{N_{2,{\rm max}}}\right)\right]n^{\rm ph}\right\}
\end{align}
and $\partial_t N_2=-\partial_t N_1$, where 
\begin{equation}
W_{12} \equiv \dfrac{2D^2}{\hbar\Delta_E}.   \label{eq:W_12}
\end{equation} 
If $N_i \gg 1$, so that $N_i \pm 1 \approx N_i$, this yields
\begin{align}
\partial_t N_1 = W_{12} \left\{N_2\left(1 - \dfrac{N_2}{N_{2,{\rm max}}}\right) \left[1 + N_1\left(1- \dfrac{N_1}{N_{1,{\rm max}}} \right) \right] \left(1+n^{\rm ph}\right)- N_1\left(1-\dfrac{N_1}{N_{1,{\rm max}}}\right)\left[1+N_2\left(1-\dfrac{N_2}{N_{2,{\rm max}}}\right)\right]n^{\rm ph}\right\},
\end{align}

In the bosonic imit, one puts $N_{i,{\rm max}} \to \infty$, and reaches the standard rate equations with bosonic stimulation term
\begin{align}\label{eq:boson_ke}
   \!\! \partial_t N_1\! =\! W_{12} \left[ N_{2} (1 + N_{1}) (1 + n^{\rm ph}) - N_{1} (1 + N_{2}) n^{\rm ph} \right].
\end{align}

In the opposite limit $N_{i,{\rm max}} = 1$ one gets the following equation:
\begin{equation}
\partial_t N_{1} = W_{12} \left[ N_{2} (2 - N_{2}) (1 - N_{1}^2) (1 + n^{\rm ph})\right. - \left.N_{1} (2 - N_{1}) (1 - N_{2}^2) n^{\rm ph} \right].\label{eqn:paulions_eq}
\end{equation}
Note, that although it contains the factors accounting for the Pauli exclusion principle, it differs from the standard fermionic kinetic equation
\begin{align}
   \!\! \partial_t N_{1}\! =\! W_{12} \left[ N_{2} (1 - N_{1}) (1 + n^{\rm ph}) - N_{1} (1 - N_{2}) n^{\rm ph} \right].\label{eq:femionic_ke}
\end{align}

Eq.~\eqref{eqn:paulions_eq} describes the dynamics of the so-called paulions, which were first introduced by Agronovich in his seminal paper describing nonlinear properties of Frenkel excitons \cite{Agranovich1968}. Paulionic operators combine the algebraic properties of bosons and fermions: they commute for the case of two different states as bosonic operators do, but anticommute for the same state as fermionic operators, which means that similar to the case of fermions one can not put more then one paulion to a given quantum state. 

It follows from Eq.~\eqref{eqn:paulions_eq} that in the linear regime ($N_{1}$, $N_{2}$ $\ll 1$), Eq.~\eqref{eqn:paulions_eq} it contains a total factor $2$, which means that spontaneous relaxation rate for paulions is twice bigger, then for elementary bosons and fermions. 

In the opposite limit $N_1$, $N_2$ $\lesssim 1$ Eq.~\eqref{eqn:paulions_eq} turns to
\begin{equation}
\partial_t N_{1} \approx 2W_{12} \left[ N_{2} (1 - N_{1}) (1 + n^{\rm ph})- N_{1}  (1 - N_{2}) n^{\rm ph} \right],\label{eqn:paulions_eq_f}
\end{equation}
which is nothing but a fermionic rate equation ~\eqref{eq:femionic_ke} with twice bigger scattering rate. 

In the case of multiple coupled energy levels corresponding to different values of momentum $\bm{p}$, the kinetic equations are straightforwardly generalize, and the corresponding system reads (see Appendix~\ref{sec:app_full_system} for the details of the derivation):
\begin{align}
    \partial_t N_{\bm{p}}=&\!\!\!\!\!\!\!\sum\limits_{\bm{q},\varepsilon_{\bm{p}+\bm{q}}>\varepsilon_{\bm{p}}}\!\!\!\!\!\!\!W_{\bm{p},\bm{p}+\bm{q}}\left\{N_{\bm{p}+\bm{q}}\left(1-\dfrac{N_{\bm{p}+\bm{q}}-1}{N_{\bm{p}+\bm{q},{\rm max}}}\right)\left[1+N_{\bm{p}}\left(1-\dfrac{N_{\bm{p}}+1}{N_{\bm{p},{\rm max}}}\right)\right]\left(1+n^{\rm ph}_{\bm{q}}\right)\right.\nonumber\\
    &\qquad\qquad\qquad\qquad\qquad\qquad\qquad\qquad\qquad\left.-N_{\bm{p}}\left(1-\dfrac{N_{\bm{p}}-1}{N_{\bm{p},{\rm max}}}\right)\left[1+N_{\bm{p}+\bm{q}}\left(1-\dfrac{N_{\bm{p}+\bm{q}}+1}{N_{\bm{p}+\bm{q},{\rm max}}}\right)\right]n^{\rm ph}_{\bm{q}}\right\}\nonumber\\
    -&\!\!\!\!\!\!\!\sum\limits_{\bm{q},\varepsilon_{\bm{p}+\bm{q}}<\varepsilon_{\bm{p}}}\!\!\!\!\!\!\!W_{\bm{p}+\bm{q},\bm{p}}\left\{N_{\bm{p}}\left(1-\dfrac{N_{\bm{p}}-1}{N_{\bm{p},{\rm max}}}\right)\left[1+N_{\bm{p}+\bm{q}}\left(1-\dfrac{N_{\bm{p}+\bm{q}}+1}{N_{\bm{p}+\bm{q},{\rm max}}}\right)\right]\left(1+n^{\rm ph}_{-\bm{q}}\right)\right.\nonumber\\
    &\qquad\qquad\qquad\qquad\qquad\qquad\qquad\qquad\qquad\left.-N_{\bm{p}+\bm{q}}\left(1-\dfrac{N_{\bm{p}+\bm{q}}-1}{N_{\bm{p}+\bm{q},{\rm max}}}\right)\left[1+N_{\bm{p}}\left(1-\dfrac{N_{\bm{p}}+1}{N_{\bm{p},{\rm max}}}\right)\right]n^{\rm ph}_{-\bm{q}}\right\}.\label{eqn:final_many_levels}
\end{align}    
\end{widetext}

\section{Numerical results and discussion \label{sec:RnD}}
Similar to the presentation of the model above, we start with the two-level system and then consider the thin two-dimensional strip.
\begin{figure}[b!]
    \centering
  \includegraphics[width=1\linewidth]{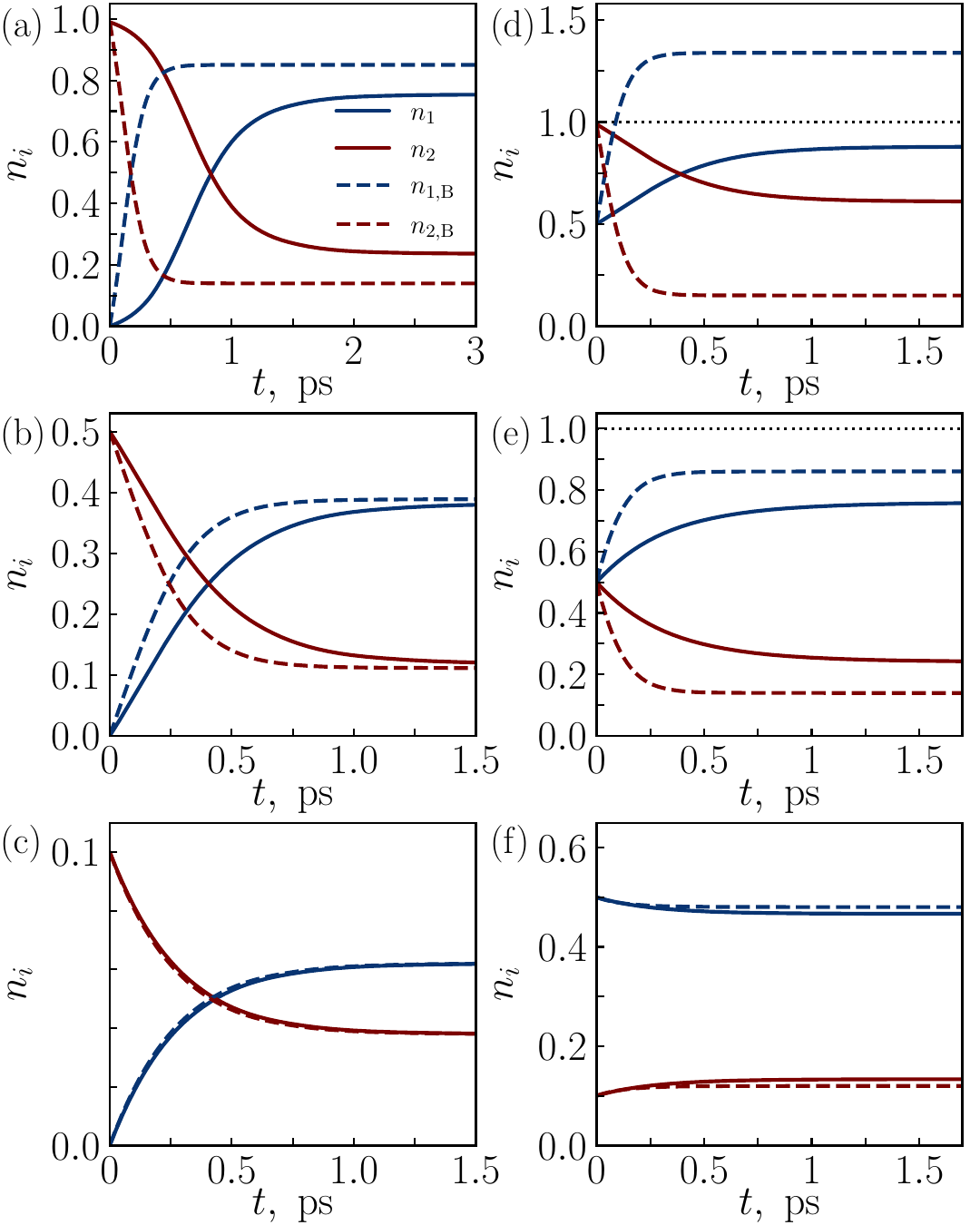}
    \caption{Temporal evolution of the relative occupation numbers $n_i=N_i/N_{i,{\rm max}}$ ($i = 1$, $2$) of a two-level system (solid lines), where $N_i$ obey Eq.~\eqref{eqn:final_two_levels}  and  their counterparts $n_{i,\textup{B}}=N_{i,\textup{B}}/N_{i,{\rm max}}$ ($i = 1$, $2$) of the bosonic limit (dashed lines), where $N_{i,\textup{B}}$ is described by Eq.~\eqref{eq:boson_ke}. For the bosonic case, the relative units of occupations, $n_{i,\textup{B}}$, are purely formal and are not required to have values limited to unity. Different panels  correspond  to different initial values of $n_i$ and $n_{i,\textup{B}}$. For panels (a),  (b), and (c), the initial values are $n_1=n_{1,\textup{B}}= 0$, while $n_2$  and $n_{2,\textup{B}}$ are assumed to equal $0.99$, $0.5$, and $0.1$, respectively. For panels (d), (e), and (f), $n_1=n_{1,\textup{B}}=0.5$ while $n_2=n_{2,\textup{B}}$ and equal to $0.99$, $0.5$, and $0.1$, respectively. The parameters chosen are $T=4~\textup{K}$, $W_{12}=0.5~\textup{ps}^{-1}$, $N_{1,{\rm max}}=N_{2,{\rm max}}=20$, and $\varepsilon^{\rm ph}=1~\textup{meV}$.}
    \label{fig:g_as_n}
\end{figure}
\subsection{Two-level system}

First, we analyze the statistical effects by comparing our predictions with those obtained in the bosonic limit in the case of a two-level system. The results of our numerical simulations are presented in Fig.~\ref{fig:g_as_n} in terms of the relative occupation numbers $n_i=N_i/N_{i,{\rm max}}$ and  $n_{i,\textup{B}}=N_{i,\textup{B}}/N_{i,{\rm max}}$ ($i = 1$, $2$), where $N_i$  and $N_{i,\textup{B}}$ satisfy Eqs.~\eqref{eqn:final_two_levels} and~\eqref{eq:boson_ke},  respectively. For the bosonic case, the relative units of occupations, $n_{i,\textup{B}}$, are purely formal and are not required to have values limited to unity, i.e., in the bosonic case for convenience of comparison and universality of behavior of curves (which would minimally depend on the maximum occupation numbers), the occupations are just measured in the units of $N_{i,{\rm max}}$. For simplicity, the maximum occupation numbers for each level are chosen to be the same $N_{1,{\rm max}}=N_{2,{\rm max}}$. In Fig.~\ref{fig:g_as_n}, we observe that, first, the total number of particles is conserved as it should be in all cases. Second, when the total starting number of particles does not exceed the maximum value of each of the levels (for panels (a), (b), (c), (e), (f), we have $n_1+n_2\leq 1$), the temporal evolution of the occupation numbers $n_i$ is not very different from the bosonic case $n_{i,\textup{B}}$ since the Pauli factors are almost irrelevant in this case. In Fig.~\ref{fig:g_as_n}, we see that only in the panel~(d) the statistical effects are important as in this case, the initial occupation of the levels yields in total $n_1+n_2=1.5$ exceeding unity. From a quantitative point of view, we find that Pauli blocking affects the relaxation rate of the system if we get closer to the occupation limit ($n_i=1$), in particular for a low total occupation number (see panels (c) and (f) in Fig.~\ref{fig:g_as_n}), the dynamics becomes identical in the both cases.

\begin{figure}[t]
    \centering
  \includegraphics[width=1\linewidth]{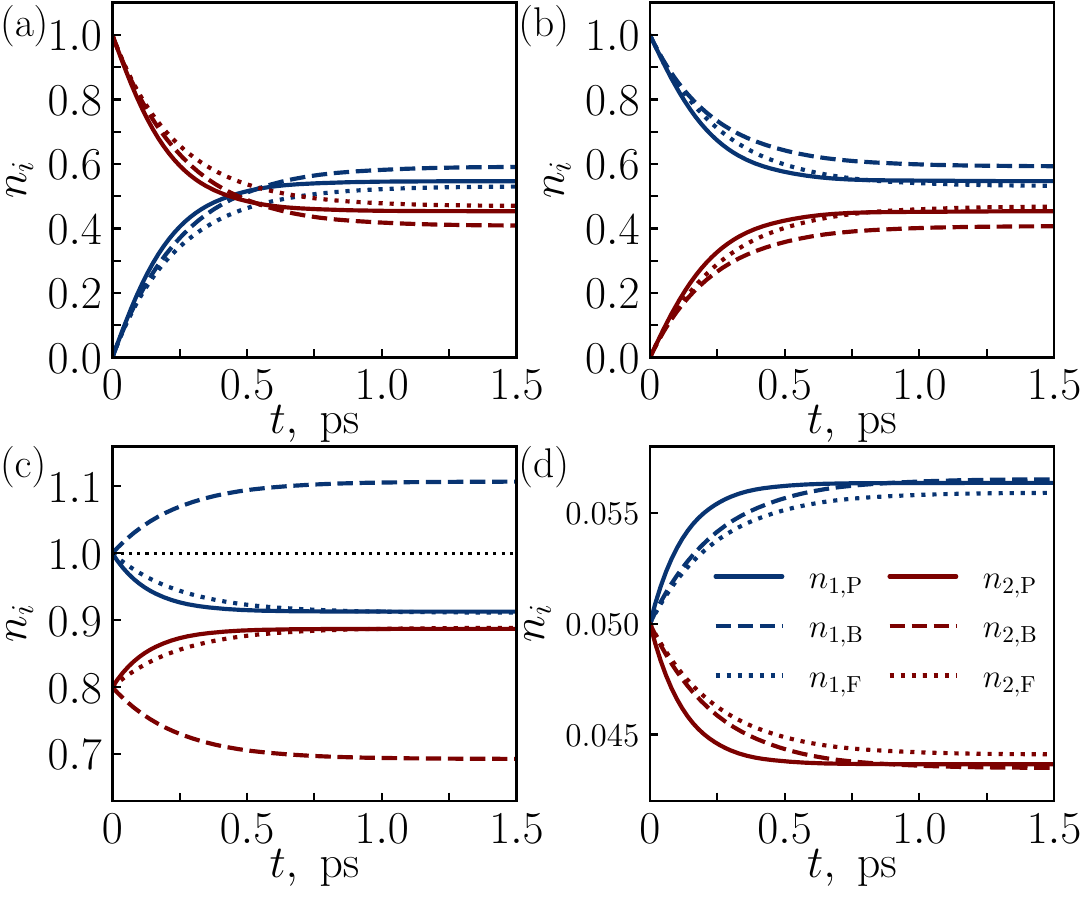}
    \caption{Temporal evolution of the relative occupation numbers $n_{i,\textup{P}}=N_{i,\textup{P}}/N_{i,{\rm max}}=N_{i,\textup{P}}$ ($i = 1$, $2$) of a two-level system in the paulionic limit (solid lines) described by Eq.~\eqref{eqn:paulions_eq}. For comparison, the bosonic [$n_{i,\textup{B}}$, dashed lines, Eq.~\eqref{eq:boson_ke}] and fermionic [$n_{i,\textup{F}}$, dotted lines, Eq.~\eqref{eq:femionic_ke}] cases are also presented. For panels (a) and (b), the initial values read $n_1=0$, $n_2 = 1$ and $n_1 = 1$, $n_2 = 0$, respectively. Panel~(c) demonstrates a difference in the behavior of the bosonic case on the one hand, and fermionic and paulionic cases on the other hand. It is evident that the fermionic and paulionic cases have the same asymptotics, only the slopes of the curves at the initial time instant differ approximately by a factor of $2$. In the opposite case of small occupancies [panel (d)], we observe that all the equations behave similarly, but the asymptotic value of the paulionic case is closer to the bosonic one. As for the slopes at the initial time instant, for the paulionic case it is again twice larger than for the others. The parameters are $T=4~\textup{K}$, $W_{12}=0.5~\textup{ps}^{-1}$, and $\varepsilon^{\rm ph}=1~\textup{meV}$.
    }
    \label{fig:g_as_n_2}
\end{figure}

In addition, we consider the limit of paulions, when there cannot be more than one particle at each level ($N_\text{max} = 1$, $n_i = N_i$). The corresponding numerical results are displayed in Fig.~\ref{fig:g_as_n_2}. We also compare them with the predictions of the bosonic [Eq.~\eqref{eq:boson_ke}] and fermionic [Eq.~\eqref{eq:femionic_ke}] kinetic equations. Here the different panels correspond again to different initial values of the occupation numbers, which are chosen to be the same for all of the equations within one panel. We observe that in the case of a small average initial occupation [panel (d)], we obtain similar results. However, we see that the asymptotic behavior of the occupations in the paulionic case is closer to their bosonic counterparts than to the fermion ones, which is natural for small occupations. On the other hand, in the case of large occupations  [see panel (c)], we reveal that the bosonic case is completely different from the others, which is natural since there is no Pauli blocking factor in the bosonic equations. Thus, for large occupation numbers, paulions behave like fermions with a relaxation rate which differs by a factor of two.

\subsection{Narrow channel}

We consider now a narrow two-dimensional strip of gallium arsenide. The exciton dispersion relation and cutoff momentum are given by
\begin{align}
\varepsilon_{\bm{k}}=E_g - E_b + \dfrac{\hbar^2 k^2}{2\mu}, \quad k_{\rm max}= \sqrt{\dfrac{2\mu E_b}{\hbar^2}},
\end{align}
where $\mu$ is the reduced mass of the electron-hole pair, $E_g$ and $E_b$ are the band gap and binding energy, respectively. The dispersion relation of the acoustic phonons reads
\begin{align}
    \varepsilon_{\bm{k}}^{\rm ph}= \hbar k v_s,
\end{align}
where $v_s$ is the speed of sound. The phonon thermal occupation is determined by the Bose distribution function according to
\begin{align}
    n_{\bm{k}}^{\rm ph}=\dfrac{1}{1-e^{\varepsilon^{\rm ph}_{\bm{k}}/(k_{\rm B}T)}}.
\end{align}
In what follows, we can vary $T$ to examine the temperature effects.

It remains to define the transition rate constant $W_{p,p+q}$. To simplify the subsequent calculations, let us focus on the one-dimensional (narrow two-dimensional channel, where the transverse momentum is neglected) setup. Although the general formalism proposed in Sec.~\ref{sec:Theory} can be applied to a full-fledged two-dimensional case, our examples will be sufficient to demonstrate the main patterns.

For the multilevel case, we utilize the following expression for the transition rate involving different wave vectors, which can be obtained by Fermi's golden rule~\cite{KIRA2006155}:
\begin{align}\label{eq:specific_w_1d_0d}
    W_{p,p+q}=|q| \alpha \delta_{\Delta_E} (\varepsilon_{p+q}-\varepsilon_{p}-\varepsilon_{q}^{\rm ph}),
\end{align}
where $\alpha$ is estimated via different material characteristics. In particular, one can resort to the formula $\alpha=(D_c-D_v)^2/\rho_m V  v_s$ (see, e.g.~\cite{HaugKoch2009}), where $\rho_m$ is the material mass density, $V$ is the structure volume, and $D_c$ and $D_v$ are the deformation constants of the conduction and valence bands, respectively. Instead, in our calculations we use some reasonable values, to avoid problems with determining the volume, mass, etc. of the sample. This is enough to demonstrate the main effect. Also, we assume that the energy profile $\delta_{\Delta_E} (\omega)$ is of the Lorentzian form:
\begin{equation}
\delta_{\Delta_E} (\omega) = \frac{\Delta_E}{\Delta_E^2 + \omega^2}.
\end{equation}
We assume further that strip length is $L_x = \SI{5}{\micro\metre}$ and the width amounts to $L_y = \SI{0.1}{\micro\metre}$. For GaAs the electron and hole effective masses are $0.067 m_0$ and $0.51 m_0$, respectively, which yields for the reduced mass $\mu=0.059 m_0$ ($m_0$ is the free electron mass). The binding energy in the two-dimensional case is $E_b=16.8~\textup{meV}$. The longitudinal sound speed is $v_s=4.73\times 10^3~\textup{m/s}$~\cite{Madelung2004}. The Bohr radius of the exciton amounts to $a_{\rm B}=2 \varepsilon_0 \varepsilon_r \hbar^2/(\mu e^2)=5.78~\textup{nm}$. Regarding the rate constant $W$, in our practical calculations instead of Eq.~\eqref{eq:specific_w_1d_0d} we used the following expression:
\begin{align}
    W_{ij}=\dfrac{W_0|j-i|}{1+\left[(\varepsilon_{k_j}-\varepsilon_{k_i}-\varepsilon^{\rm ph}_{k_{j-i}})/\Delta_E\right]^2},
\end{align}
where the discrete indices $i$ and $j$ vary from $1$ to $k_{\rm max}/\Delta k=398$   (total number of levels). Here $k_{\rm max}$ is defined via $\varepsilon_{k_{\rm max}}=E_g$, while $\Delta k=2\pi/L_x$ according to the standard box discretization. In our calculations, we set $W_0=0.005~\textup{ps}^{-1}$. In line with Eq.~\eqref{eq:N0max}, the maximum occupancy of the level is $N_{i,{\rm max}}=5L_xL_y/(4\pi a_{\rm B}^2) = 5.9\times 10^3$ for all $i$. Note that in terms of the density, it yields $n_{\text{X,max}} \approx 1.2\times 10^{12}~\textup{cm}^{-2}$, which one may have expected based on the qualitative condition~\eqref{eq:condition}.

\begin{figure}[t]
    \centering
\includegraphics[width=1\linewidth]{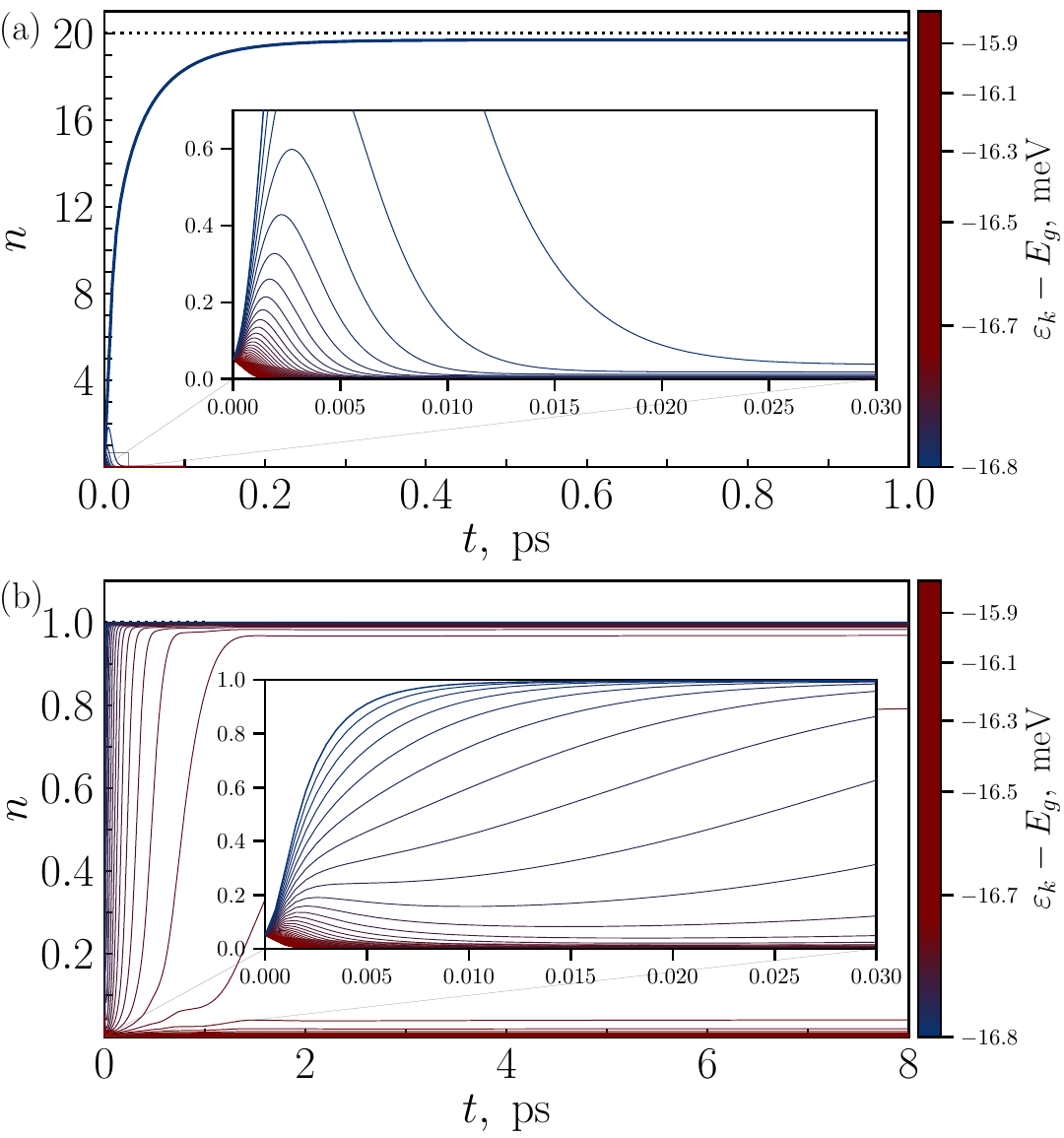}
\caption{Temporal evolution of the relative occupation numbers $n_i=N_i/N_{i,{\rm max}}$ of a multilevel narrow-channel system obtained in the bosonic limit~(a) and revealed by means of our full numerical simulations according to Eqs.~\eqref{eqn:final_many_levels}~(b). The initial values are $n_i (t=0) = 0.05$. The curves are color-coded according to the energies displayed in the color bar. The temperature is $T=10~\textup{K}$. We display only the $100$ lowest-energy levels.}
\label{fig:g_as_n_3}
\end{figure}

We solve the resulting system of equations~\eqref{eqn:final_many_levels} for $T=10~\textup{K}$ and two different initial occupations: $n_i = 0.05$ and  $n_i = 0.004$. These initial values are chosen in order to clearly demonstrate the similarities and differences between the full actual behavior of the cobosons and its bosonic limit. First, we depict the temporal evolution of the occupation numbers for $n_{i}(t=0)=0.05$ (see Fig.~\ref{fig:g_as_n_3}). In panels~(a) and~(b), we display the bosonic limit and the predictions of the full system~\eqref{eqn:final_many_levels}, respectively.
\begin{figure}[t]
    \centering
\includegraphics[width=1\linewidth]{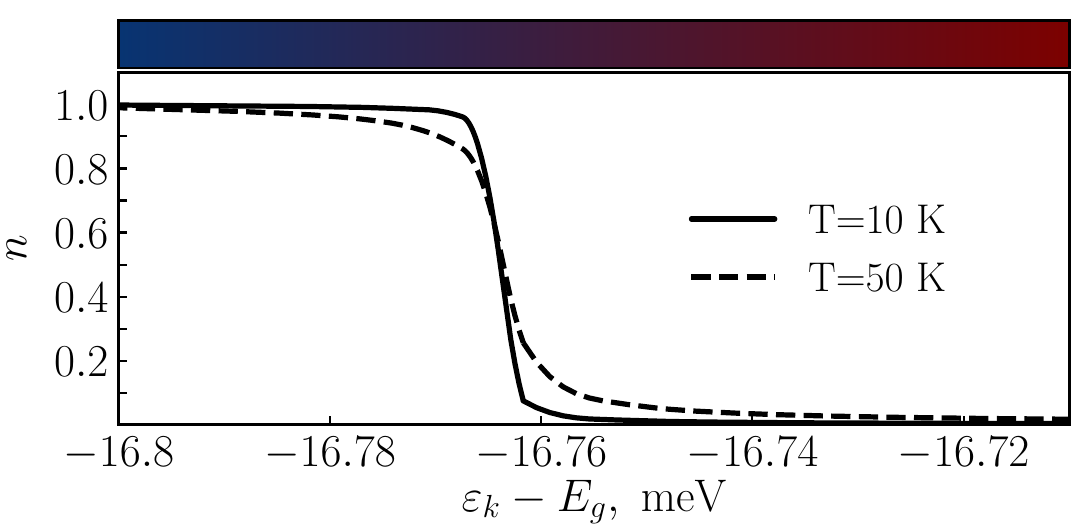}
\caption{Final (asymptotic) values of the relative occupation numbers $n_i=N_i/N_{i,{\rm max}}$ as a function of energy obtained in the coboson case by our numerical computations according to the kinetic equations~\eqref{eqn:final_many_levels} for two different values of the temperature. The left edge of the energy spectrum corresponds to $\varepsilon_k - E_g = -E_b$, where $E_b=16.8~\textup{meV}$ is the binding energy. The resulting distributions resemble the Fermi-Dirac ones. The initial occupations in both cases are $N_i = 0.05 N_{i,{\rm max}}$, and the number of levels is $398$.}
\label{fig:g_as_n_4}
\end{figure}
In the bosonic limit, all of the particles plainly fall into the ground state, so that the corresponding relative occupation number, which has here a purely formal definition, tends to the value which can be estimated as a product of the initial relative occupation and the number of levels: $0.05 \times 398 \approx 20$. On the other hand, in the case of our full-scale simulations based on Eqs.~\eqref{eqn:final_many_levels} and taking into account the nonbosonic nature of the excitons, the maximum relative occupation number is exactly $1$. In Fig.~\ref{fig:g_as_n_3}(b) we observe the following redistribution mechanism: the particles fully occupy the energy levels one by one starting from the ground state. This process lasts until $n_i$ reaches unity for approximately 20 lowest levels, so that all of the particles reside in these levels (due to a nonzero temperature, $n_i$ changes smoothly from $1$ to $0$ in the vicinity of $i=20$). We also note that the composite structure of the excitons affects not only the maximum occupation number, but also the relaxation rate. To address the temperature effects, we demonstrate the final (stationary) occupation numbers as a function of energy for two different values of $T$ (see Fig.~\ref{fig:g_as_n_4}). Here the initial occupation numbers remain the same [$n_{i}(t=0)=0.05$], and we display only the coboson case since in the bosonic one, the structure is trivial: it contains only a sharp peak corresponding to the ground state. Note that the stationary picture revealed here proves to have a form which resembles, though does not exactly reproduce, the Fermi-Dirac distribution (with a maximum occupation number $N_{i,{\rm max}}$ instead of $1$). A thorough analysis of the properties of this distribution represents an important task, which we would like to tackle in the future.

\begin{figure}[t]
    \centering
\includegraphics[width=1\linewidth]{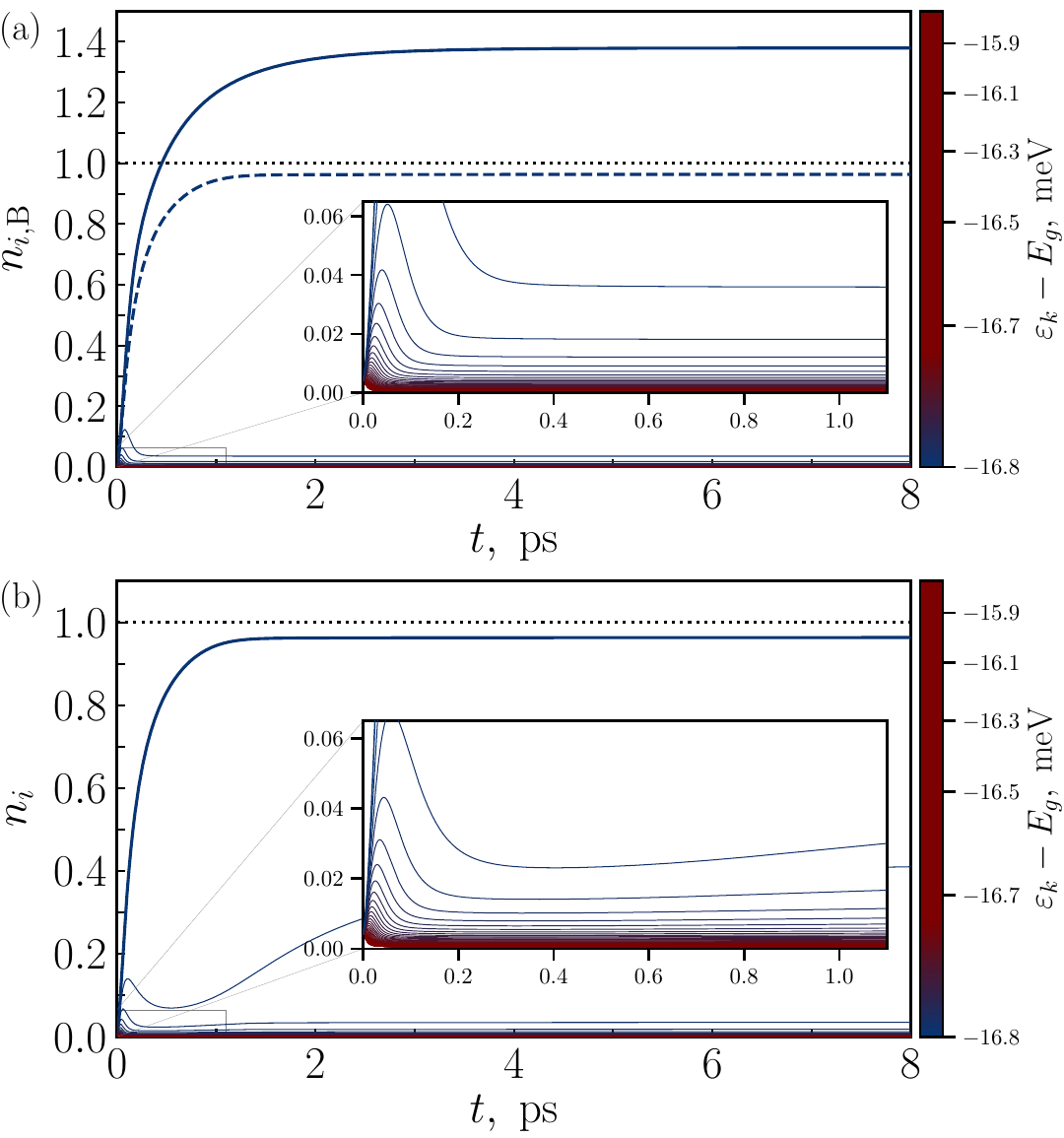}
\caption{Temporal evolution of the relative occupation numbers $n_i=N_i/N_{i,{\rm max}}$ of a multilevel narrow-channel system obtained in the bosonic limit~(a) and found via the full system~\eqref{eqn:final_many_levels}~(b). The initial values are $n_i (t=0) = 0.004$. The dashed blue line in the panel~(a) corresponds to the cobosonic occupation of the ground state. In the both panels, the black dotted line shows the maximal occupation number in the case of cobosons ($n_i=1$). The solid curves are color-coded according to the energies displayed in the color bar. The temperature is $T=10~\textup{K}$. We display only the $100$ lowest-energy levels.}
\label{fig:g_as_n_5}
\end{figure}

Finally, in Fig.~\ref{fig:g_as_n_5} we consider the opposite scenario, where the initial occupation numbers are very small [$n_{i}(t=0)=0.004$]. In this case, the numerical predictions obtained via the full system~\eqref{eqn:final_many_levels} essentially coincide with the bosonic limit for all $i \geq 2$. Nevertheless, the relative occupation of the ground state in the bosonic limit exhibits a different behavior as it exceeds $1$ at a certain point, which is prohibited in the case of cobosons. Since the total number of particles is small, the excited states ($i \geq 2$) possess very small occupation numbers in the both panels and the statistical properties are irrelevant here (there is, again, a smearing effect due to the nonzero temperature).

Our results are readily comprehensible from the physical viewpoint, which evidently suggests that the proposed model is adequate and self-consistent. We emphasize that the key application of the kinetic equations derived in our study concerns a quantitative and qualitative description of dynamical relaxation processes in complex many-particle systems.

\section{Conclusion \label{sec:Conclusion}}

We developed a theoretical model describing exciton-phonon relaxation dynamics accounting for the composite structure of the excitons. We discussed the problem of the proper definition of excitonic creation and annihilation operators based angular momentum algebra and derived modified semiclassical Boltzmann equations which capture the corresponding statistical effects.

Our analysis demonstrates that the composite nature of the excitons significantly alters the kinetic equations governing their dynamics. The inclusion of statistical factors reflecting the finite maximum occupancy of the exciton states results in deviations from the ideal bosonic behavior, especially at high exciton densities. By performing numerical simulations based on our equations, we showed that the thermalization processes are influenced by the nonbosonic properties of the excitons, which affects observable quantities such as the final energy distributions and relaxation times.

Our findings highlight the importance of considering the internal structure and statistical properties of excitons in theoretical models, particularly for systems operating in regimes where the exciton densities are high. Future research may extend the developed approach and include additional types of the interactions, such as exciton-exciton scattering, providing a more comprehensive understanding of the dynamics of complex excitonic systems.

\section{Acknowledgements}
A.K. acknowledges the support of the Icelandic Research Fund (Ranns\'oknasj\'o{\dh}ur, Grant No.~2410550) and the Ministry of Science and Higher Education of the Russian Federation (Goszadaniye) Project No. FSMG-2023-0011.

\begin{appendix}

\subsection*{Appendix: Derivation of the kinetic equations in the case of multiple energy levels \label{sec:app_full_system}}

The Hamiltonian of our model in the case of an arbitrary number of the energy levels within the interaction picture reads
\begin{multline}
    \hat{H} = \sum\limits_{\bm{k}} \varepsilon_{\bm{k}} \hat{N}_{\bm{k}} + \sum\limits_{\bm{q}} \varepsilon_{\bm{q}}^{\rm ph} \, \hat{b}^{\dagger}_{\bm{q}} \hat{b}_{\bm{q}} \\
+\sum\limits_{\bm{k},\bm{q}} \tilde{D}(\bm{q}) \left[ \hat{J}_{\bm{k}+\bm{q}}^{+} \hat{J}_{\bm{k}}^{-} \hat{b}_{\bm{q}} e^{{\rm i} (\varepsilon_{\bm{k}+\bm{q}} - \varepsilon_{\bm{k}} - \varepsilon_{\bm{q}}^{\rm ph}) t/\hbar} \right. \\
    \left. + \hat{J}_{\bm{k}}^{+} \hat{J}_{\bm{k}+\bm{q}}^{-} \hat{b}_{\bm{q}}^{\dagger} e^{-{\rm i} (\varepsilon_{\bm{k}+\bm{q}} - \varepsilon_{\bm{k}} - \varepsilon_{\bm{q}}^{\rm ph}) t/\hbar} \right],
\end{multline}
where $\hat{J}_{\bm{k}}^{\pm}$ are the raising and lowering operators for the excitons in state $\bm{k}$, and $\hat{N}_{\bm{k}}$ is the number (density) operator for state $\bm{k}$.
To handle the interaction with the phonon reservoir, we represent the Hamiltonian as $\hat{H} = \hat{H}_{\rm S} + \hat{H}_{\rm R} + \hat{H}_{\rm SR}$, where $\hat{H}_{\rm S} = \sum_{\bm{k}} \varepsilon_{\bm{k}} \hat{N}_{\bm{k}}$ is the free exciton Hamiltonian, $\hat{H}_{\rm R} = \sum_{\bm{q}} \varepsilon_{\bm{q}}^{\rm ph} \, \hat{b}^{\dagger}_{\bm{q}} \hat{b}_{\bm{q}}$ is the reservoir Hamiltonian, and $\hat{H}_{\rm SR}$ is the exciton-reservoir interaction Hamiltonian. 
The evolution equation (the Liouville--von Neumann equation for the density matrix) takes the form
\begin{equation}
    {\rm i} \hbar \, \partial_t \hat{\chi}(t) = \big [ \hat{H}_{\rm SR}(t), \hat{\chi}(t) \big ],
\end{equation}
which we formally integrate and obtain
\begin{equation}
    \hat{\chi}(t) = \hat{\chi}(-\infty) + \frac{1}{{\rm i} \hbar} \int\limits_{-\infty}^{t} \! {\rm d}t' \big[ \hat{H}_{\rm SR}(t'), \hat{\chi}(t') \big].
\end{equation}
Assuming that $\hat{\chi}(-\infty)$ commutes with $\hat{H}_{\rm SR}$, we find
\begin{equation}
    \partial_t \hat{\chi}(t) = -\frac{1}{\hbar^2} \int\limits_{-\infty}^{t} \! {\rm d}t' \big[ \hat{H}_{\rm SR}(t), \big[ \hat{H}_{\rm SR}(t'), \hat{\chi}(t') \big] \big].
\end{equation}
By applying now the Born-Markov approximation $\hat{\chi}(t') \to \hat{\chi}(t)$, we obtain
\begin{equation}
    \partial_t \hat{\chi}(t) = -\frac{1}{\hbar^2} \int\limits_{-\infty}^{t} \! {\rm d}t' \big[ \hat{H}_{\rm SR}(t), \big[ \hat{H}_{\rm SR}(t'), \hat{\chi}(t) \big] \big].
\end{equation}
The time integral yields the functions $\delta_{\Delta_E}$ imposing the energy conservation:
\begin{align}
\dfrac{1}{\hbar}\int\limits_{-\infty}^{t} \!\! {\rm d}t'e^{{\rm i}(\varepsilon_{\bm{k}+\bm{q}}-\varepsilon_{\bm{k}}-\varepsilon^{\rm  ph}_{\bm{q}})t'/\hbar} &\approx \delta_{\Delta_E} (\varepsilon_{\bm{k}+\bm{q}}-\varepsilon_{\bm{k}}-\varepsilon^{\rm  ph}_{\bm{q}}).
\end{align}
This simplifies the evolution equation:
\begin{equation}\label{app:eqn:dens_matr_final_eq}
   \partial_t\hat{\chi}=\dfrac{1}{\hbar}\left(\comm\big{\hat{H}^+}{\comm\big{\hat{\chi}}{\hat{h}^-}}+\comm\big{\hat{H}^-}{\comm\big{\hat{\chi}}{\hat{h}^+}}\right),
\end{equation}
where
\begin{align}
&\hat{H}^+ = \sum_{\bm{k}, \bm{q}} \tilde{D}(\bm{q}) \hat{J}^+_{\bm{k}} \hat{J}^-_{\bm{k}+\bm{q}} \hat{b}_{\bm{q}}^{\dagger} e^{-{\rm i} (\varepsilon_{\bm{k}+\bm{q}}-\varepsilon_{\bm{k}}-\varepsilon^{\rm ph}_{\bm{q}}) t/\hbar},\\
&\hat{H}^- = \sum_{\bm{k}, \bm{q}} \tilde{D}(q) \hat{J}^+_{\bm{k}+\bm{q}} \hat{J}^-_{\bm{k}} \hat{b}_{\bm{q}} e^{{\rm i}(\varepsilon_{\bm{k}+\bm{q}}-\varepsilon_{\bm{k}}-\varepsilon^{\rm ph}_{\bm{q}}) t/\hbar},\\
&\hat{h}^- =\!\!\!\!\!\! \sum_{\substack{\bm{k}, \bm{q} \\ \varepsilon_{\bm{k}+\bm{q}} > \varepsilon_{\bm{k}}}}\!\!\!\! \tilde{D}(\bm{q}) \hat{J}^+_{\bm{k}+\bm{q}} \hat{J}^-_{\bm{k}} \hat{b}_{\bm{q}} \delta_{\Delta_E}(\varepsilon_{\bm{k}+\bm{q}} - \varepsilon_{\bm{k}} - \varepsilon^{\rm ph}_{\bm{q}}),\\
&\hat{h}^+ =\!\!\!\!\!\!  \sum_{\substack{\bm{k}, \bm{q} \\ \varepsilon_{\bm{k}} > \varepsilon_{\bm{k}+\bm{q}}}}\!\!\!\!  \tilde{D}(\bm{q}) \hat{J}^+_{\bm{k}+\bm{q}} \hat{J}^-_{\bm{k}} \hat{b}^\dagger_{-\bm{q}} \delta_{\Delta_E}(\varepsilon_{\bm{k}} - \varepsilon_{\bm{k}+\bm{q}} - \varepsilon^{\rm ph}_{\bm{q}}).\!\!\!\!
\end{align}
Just as was done in the case of a two-level system [see Eq.~\eqref{eq:motion_of_Jz}], we reduce the problem to the following equation:
\begin{equation}
    \partial_t  \big \langle \hat{J}^z_{\bm{p}} \big\rangle = \frac{2}{\hbar} \, \text{Re} \, \big\langle \comm\big{\hat{h}^-}{\comm\big{\hat{J}_{\bm{p}}^z}{\hat{H}^+}} \big\rangle.
\end{equation}
Accordingly, we evaluate the commutators
\begin{align}
\big[ \hat{J}_{\bm{p}}^z, \hat{H}^+ \big] &= \sum\limits_{\bm{q}} \tilde{D}(\bm{q}) e^{{\rm i} \varepsilon^{\rm ph}_{\bm{q}} t/\hbar}\hat{b}_{\bm{q}}^\dagger \left[ \hat{J}_{\bm{p}}^+ \hat{J}_{\bm{p}+\bm{q}}^- e^{{\rm i} (\varepsilon_{\bm{p}} - \varepsilon_{\bm{p}+\bm{q}}) t/\hbar}\right.\nonumber \\
&-\left.\hat{J}_{\bm{p}-\bm{q}}^+ \hat{J}_{\bm{p}}^- e^{{\rm i} (\varepsilon_{\bm{p}-\bm{q}} - \varepsilon_{\bm{p}}) t/\hbar} \right]
\end{align}
and then
\begin{align}
&\big[ \hat{h}^-, \big[\hat{J}_{\bm{p}}^z, H^+ \big]\big] = \!\!\!\!\!\!\! \sum_{\bm{q},\varepsilon_{\bm{p}+\bm{q}} > \varepsilon_{\bm{p}}} \!\!\!\!\!\!\! \tilde{D}^2 (\bm{q}) \delta_{\Delta_E}(\varepsilon_{\bm{p}+\bm{q}} - \varepsilon_{\bm{p}} - \varepsilon^{\rm ph}_{\bm{q}}) \nonumber \\
{}&\quad\times \big ( \hat{J}^{+}_{\bm{p}+\bm{q}} \hat{J}^{-}_{\bm{p}+\bm{q}} \hat{J}^{-}_{\bm{p}} \hat{J}^{+}_{\bm{p}} \hat{b}_{\bm{q}} \hat{b}_{\bm{q}}^\dagger - \hat{J}^{-}_{\bm{p}+\bm{q}} \hat{J}^{+}_{\bm{p}+\bm{q}} \hat{J}^{+}_{\bm{p}} \hat{J}^{-}_{\bm{p}} \hat{b}_{\bm{q}}^\dagger \hat{b}_{\bm{q}} \big ) \nonumber \\
&\qquad\qquad\qquad {}- \!\!\!\!\!\! \sum_{\bm{q},\varepsilon_{\bm{p}+\bm{q}} < \varepsilon_{\bm{p}}} \!\!\!\!\!\!\! \tilde{D}^2 (\bm{q}) \delta_{\Delta_E}(\varepsilon_{\bm{p}} - \varepsilon_{\bm{p}+\bm{q}} - \varepsilon^{\rm ph}_{-\bm{q}}) \nonumber \\
{}&\quad\times \big ( \hat{J}^{-}_{\bm{p}+\bm{q}} \hat{J}^{+}_{\bm{p}+\bm{q}} \hat{J}^{+}_{\bm{p}} \hat{J}^{-}_{\bm{p}} \hat{b}_{-\bm{q}} \hat{b}_{-\bm{q}}^\dagger - \hat{J}^{+}_{\bm{p}+\bm{q}} \hat{J}^{-}_{\bm{p}+\bm{q}} \hat{J}^{-}_{\bm{p}} \hat{J}^{+}_{\bm{p}} \hat{b}_{-\bm{q}}^\dagger \hat{b}_{-\bm{q}} \big ).
\end{align}
By combining the commutators and moving on to the number of particles $N_{\bm{k}} = \langle \hat{N}_{\bm{k}} \rangle $, we obtain the following equations:
\begin{align}
    \partial_t N_{\bm{p}}=&\!\!\!\!\!\!\!\sum\limits_{\bm{q},\varepsilon_{\bm{p}+\bm{q}}>\varepsilon_{\bm{p}}}\!\!\!\!\!\!\!W_{\bm{p},\bm{p}+\bm{q}}\left[\langle \hat{J}^{+}_{\bm{p}+\bm{q}}\hat{J}^{-}_{\bm{p}+\bm{q}} \rangle\langle \hat{J}^{-}_{\bm{p}}\hat{J}^{+}_{\bm{p}} \rangle \left(1+n^{\rm ph}_{\bm{q}}\right)\right.\nonumber\\
    &\qquad\qquad\left.-\langle \hat{J}^{-}_{\bm{p}+\bm{q}}\hat{J}^{+}_{\bm{p}+\bm{q}} \rangle\langle \hat{J}^{+}_{\bm{p}}\hat{J}^{-}_{\bm{p}} \rangle n^{\rm ph}_{\bm{q}}\right]\nonumber\\
    -&\!\!\!\!\!\!\!\sum\limits_{\bm{q},\varepsilon_{\bm{p}+\bm{q}}<\varepsilon_{\bm{p}}}\!\!\!\!\!\!\!W_{\bm{p}+\bm{q},\bm{p}}\left[\langle \hat{J}^{-}_{\bm{p}+\bm{q}}\hat{J}^{+}_{\bm{p}+\bm{q}} \rangle\langle \hat{J}^{+}_{\bm{p}}\hat{J}^{-}_{\bm{p}} \rangle \left(1+n^{\rm ph}_{-\bm{q}}\right)\right.\nonumber\\
    &\qquad\qquad\left.-\langle \hat{J}^{+}_{\bm{p}+\bm{q}}\hat{J}^{-}_{\bm{p}+\bm{q}} \rangle\langle \hat{J}^{-}_{\bm{p}}\hat{J}^{+}_{\bm{p}} \rangle n^{\rm ph}_{-\bm{q}}\right]. \label{eqn:app:final_many_levels}
\end{align}
The last step to take is to rewrite the approximate expressions~\eqref{eq:approx_laderpm} and~\eqref{eq:approx_ladermp}. After substituting them, we arrive at the final system~\eqref{eqn:final_many_levels}.

While deriving the equations of motions, we made similar approximations as in the two-level case. We neglected the higher-order correlations in order to factorize the expectation values. For operators like $\hat{N}_{\bm{k}}^2$, we used the approximation $\langle \hat{N}_{\bm{k}}^2 \rangle \approx N_{\bm{k}}^2$. This is justified when the fluctuations are minor.

\end{appendix}

\bibliography{bibitems}

\begin{thebibliography}{40}%
\makeatletter
\providecommand \@ifxundefined [1]{%
 \@ifx{#1\undefined}
}%
\providecommand \@ifnum [1]{%
 \ifnum #1\expandafter \@firstoftwo
 \else \expandafter \@secondoftwo
 \fi
}%
\providecommand \@ifx [1]{%
 \ifx #1\expandafter \@firstoftwo
 \else \expandafter \@secondoftwo
 \fi
}%
\providecommand \natexlab [1]{#1}%
\providecommand \enquote  [1]{``#1''}%
\providecommand \bibnamefont  [1]{#1}%
\providecommand \bibfnamefont [1]{#1}%
\providecommand \citenamefont [1]{#1}%
\providecommand \href@noop [0]{\@secondoftwo}%
\providecommand \href [0]{\begingroup \@sanitize@url \@href}%
\providecommand \@href[1]{\@@startlink{#1}\@@href}%
\providecommand \@@href[1]{\endgroup#1\@@endlink}%
\providecommand \@sanitize@url [0]{\catcode `\\12\catcode `\$12\catcode `\&12\catcode `\#12\catcode `\^12\catcode `\_12\catcode `\%12\relax}%
\providecommand \@@startlink[1]{}%
\providecommand \@@endlink[0]{}%
\providecommand \url  [0]{\begingroup\@sanitize@url \@url }%
\providecommand \@url [1]{\endgroup\@href {#1}{\urlprefix }}%
\providecommand \urlprefix  [0]{URL }%
\providecommand \Eprint [0]{\href }%
\providecommand \doibase [0]{https://doi.org/}%
\providecommand \selectlanguage [0]{\@gobble}%
\providecommand \bibinfo  [0]{\@secondoftwo}%
\providecommand \bibfield  [0]{\@secondoftwo}%
\providecommand \translation [1]{[#1]}%
\providecommand \BibitemOpen [0]{}%
\providecommand \bibitemStop [0]{}%
\providecommand \bibitemNoStop [0]{.\EOS\space}%
\providecommand \EOS [0]{\spacefactor3000\relax}%
\providecommand \BibitemShut  [1]{\csname bibitem#1\endcsname}%
\let\auto@bib@innerbib\@empty
\bibitem [{\citenamefont {Haug}\ and\ \citenamefont {Koch}(2009)}]{HaugKoch2009}%
  \BibitemOpen
  \bibfield  {author} {\bibinfo {author} {\bibfnamefont {H.}~\bibnamefont {Haug}}\ and\ \bibinfo {author} {\bibfnamefont {S.}~\bibnamefont {Koch}},\ }\href {https://doi.org/10.1142/7184} {\emph {\bibinfo {title} {Quantum Theory of the Optical and Electronic Properties of Semiconductors}}},\ \bibinfo {edition} {5th}\ ed.\ (\bibinfo  {publisher} {World Scientific},\ \bibinfo {year} {2009})\BibitemShut {NoStop}%
\bibitem [{\citenamefont {Knox}(1963)}]{Knox1963}%
  \BibitemOpen
  \bibfield  {author} {\bibinfo {author} {\bibfnamefont {R.}~\bibnamefont {Knox}},\ }\href@noop {} {\emph {\bibinfo {title} {Theory of Excitons}}}\ (\bibinfo  {publisher} {Academic Press},\ \bibinfo {year} {1963})\ \bibinfo {note} {new York and London}\BibitemShut {NoStop}%
\bibitem [{\citenamefont {Kavokin}\ \emph {et~al.}(2017)\citenamefont {Kavokin}, \citenamefont {Baumberg}, \citenamefont {Malpuech},\ and\ \citenamefont {Laussy}}]{Kavokin2017_OxfPr}%
  \BibitemOpen
  \bibfield  {author} {\bibinfo {author} {\bibfnamefont {A.}~\bibnamefont {Kavokin}}, \bibinfo {author} {\bibfnamefont {J.}~\bibnamefont {Baumberg}}, \bibinfo {author} {\bibfnamefont {G.}~\bibnamefont {Malpuech}},\ and\ \bibinfo {author} {\bibfnamefont {F.}~\bibnamefont {Laussy}},\ }\href {https://doi.org/10.1093/oso/9780198782995.001.0001} {\emph {\bibinfo {title} {Microcavities}}}\ (\bibinfo  {publisher} {Oxford University Press},\ \bibinfo {year} {2017})\BibitemShut {NoStop}%
\bibitem [{\citenamefont {Tassone}\ and\ \citenamefont {Yamamoto}(1999)}]{Tassone1999}%
  \BibitemOpen
  \bibfield  {author} {\bibinfo {author} {\bibfnamefont {F.}~\bibnamefont {Tassone}}\ and\ \bibinfo {author} {\bibfnamefont {Y.}~\bibnamefont {Yamamoto}},\ }\bibfield  {title} {\bibinfo {title} {Exciton-exciton scattering dynamics in a semiconductor microcavity and stimulated scattering into polaritons},\ }\href {https://doi.org/10.1103/PhysRevB.59.10830} {\bibfield  {journal} {\bibinfo  {journal} {Phys. Rev. B}\ }\textbf {\bibinfo {volume} {59}},\ \bibinfo {pages} {10830} (\bibinfo {year} {1999})}\BibitemShut {NoStop}%
\bibitem [{\citenamefont {Huang}\ \emph {et~al.}(2000)\citenamefont {Huang}, \citenamefont {Tassone},\ and\ \citenamefont {Yamamoto}}]{Huang2000}%
  \BibitemOpen
  \bibfield  {author} {\bibinfo {author} {\bibfnamefont {R.}~\bibnamefont {Huang}}, \bibinfo {author} {\bibfnamefont {F.}~\bibnamefont {Tassone}},\ and\ \bibinfo {author} {\bibfnamefont {Y.}~\bibnamefont {Yamamoto}},\ }\bibfield  {title} {\bibinfo {title} {Experimental evidence of stimulated scattering of excitons into microcavity polaritons},\ }\href {https://doi.org/10.1103/PhysRevB.61.R7854} {\bibfield  {journal} {\bibinfo  {journal} {Phys. Rev. B}\ }\textbf {\bibinfo {volume} {61}},\ \bibinfo {pages} {R7854} (\bibinfo {year} {2000})}\BibitemShut {NoStop}%
\bibitem [{\citenamefont {Butov}\ \emph {et~al.}(2001)\citenamefont {Butov}, \citenamefont {Ivanov}, \citenamefont {Imamoglu}, \citenamefont {Littlewood}, \citenamefont {Shashkin}, \citenamefont {Dolgopolov}, \citenamefont {Campman},\ and\ \citenamefont {Gossard}}]{Butov2001}%
  \BibitemOpen
  \bibfield  {author} {\bibinfo {author} {\bibfnamefont {L.}~\bibnamefont {Butov}}, \bibinfo {author} {\bibfnamefont {A.}~\bibnamefont {Ivanov}}, \bibinfo {author} {\bibfnamefont {A.}~\bibnamefont {Imamoglu}}, \bibinfo {author} {\bibfnamefont {P.}~\bibnamefont {Littlewood}}, \bibinfo {author} {\bibfnamefont {A.}~\bibnamefont {Shashkin}}, \bibinfo {author} {\bibfnamefont {V.}~\bibnamefont {Dolgopolov}}, \bibinfo {author} {\bibfnamefont {K.}~\bibnamefont {Campman}},\ and\ \bibinfo {author} {\bibfnamefont {A.}~\bibnamefont {Gossard}},\ }\bibfield  {title} {\bibinfo {title} {Stimulated scattering of indirect excitons in coupled quantum wells: Signature of a degenerate bose-gas of excitons},\ }\href {https://doi.org/10.1103/PhysRevLett.86.5608} {\bibfield  {journal} {\bibinfo  {journal} {Phys. Rev. Lett.}\ }\textbf {\bibinfo {volume} {86}},\ \bibinfo {pages} {5608} (\bibinfo {year} {2001})}\BibitemShut {NoStop}%
\bibitem [{\citenamefont {Butov}\ \emph {et~al.}(2002)\citenamefont {Butov}, \citenamefont {Lai}, \citenamefont {Ivanov}, \citenamefont {Gossard},\ and\ \citenamefont {Chemla}}]{Butov2002}%
  \BibitemOpen
  \bibfield  {author} {\bibinfo {author} {\bibfnamefont {L.}~\bibnamefont {Butov}}, \bibinfo {author} {\bibfnamefont {C.}~\bibnamefont {Lai}}, \bibinfo {author} {\bibfnamefont {A.}~\bibnamefont {Ivanov}}, \bibinfo {author} {\bibfnamefont {A.}~\bibnamefont {Gossard}},\ and\ \bibinfo {author} {\bibfnamefont {D.}~\bibnamefont {Chemla}},\ }\bibfield  {title} {\bibinfo {title} {Towards bose-einstein condensation of excitons in potential traps},\ }\href {https://doi.org/10.1038/417047a} {\bibfield  {journal} {\bibinfo  {journal} {Nature}\ }\textbf {\bibinfo {volume} {417}},\ \bibinfo {pages} {47} (\bibinfo {year} {2002})}\BibitemShut {NoStop}%
\bibitem [{\citenamefont {Morita}\ \emph {et~al.}(2022)\citenamefont {Morita}, \citenamefont {Yoshioka},\ and\ \citenamefont {Kuwata-Gonokami}}]{Morita2022}%
  \BibitemOpen
  \bibfield  {author} {\bibinfo {author} {\bibfnamefont {Y.}~\bibnamefont {Morita}}, \bibinfo {author} {\bibfnamefont {K.}~\bibnamefont {Yoshioka}},\ and\ \bibinfo {author} {\bibfnamefont {M.}~\bibnamefont {Kuwata-Gonokami}},\ }\bibfield  {title} {\bibinfo {title} {Observation of bose-einstein condensates of excitons in a bulk semiconductor},\ }\href {https://doi.org/10.1038/s41467-022-33103-4} {\bibfield  {journal} {\bibinfo  {journal} {Nat. Commun.}\ }\textbf {\bibinfo {volume} {13}},\ \bibinfo {pages} {5388} (\bibinfo {year} {2022})}\BibitemShut {NoStop}%
\bibitem [{\citenamefont {Gurioli}\ \emph {et~al.}(1998)\citenamefont {Gurioli}, \citenamefont {Borri}, \citenamefont {Colocci}, \citenamefont {Gulia}, \citenamefont {Rossi}, \citenamefont {Molinari}, \citenamefont {Selbmann},\ and\ \citenamefont {Lugli}}]{Gurioli1998_PRB}%
  \BibitemOpen
  \bibfield  {author} {\bibinfo {author} {\bibfnamefont {M.}~\bibnamefont {Gurioli}}, \bibinfo {author} {\bibfnamefont {P.}~\bibnamefont {Borri}}, \bibinfo {author} {\bibfnamefont {M.}~\bibnamefont {Colocci}}, \bibinfo {author} {\bibfnamefont {M.}~\bibnamefont {Gulia}}, \bibinfo {author} {\bibfnamefont {F.}~\bibnamefont {Rossi}}, \bibinfo {author} {\bibfnamefont {E.}~\bibnamefont {Molinari}}, \bibinfo {author} {\bibfnamefont {P.}~\bibnamefont {Selbmann}},\ and\ \bibinfo {author} {\bibfnamefont {P.}~\bibnamefont {Lugli}},\ }\bibfield  {title} {\bibinfo {title} {Exciton formation and relaxation in gaas epilayers},\ }\href {https://doi.org/10.1103/PhysRevB.58.R13403} {\bibfield  {journal} {\bibinfo  {journal} {Phys. Rev. B}\ }\textbf {\bibinfo {volume} {58}},\ \bibinfo {pages} {R13403} (\bibinfo {year} {1998})}\BibitemShut {NoStop}%
\bibitem [{\citenamefont {Vinattieri}\ \emph {et~al.}(1994)\citenamefont {Vinattieri}, \citenamefont {Shah}, \citenamefont {Damen}, \citenamefont {Kim}, \citenamefont {Pfeiffer}, \citenamefont {Maialle},\ and\ \citenamefont {Sham}}]{Vinattieri1994_PRB}%
  \BibitemOpen
  \bibfield  {author} {\bibinfo {author} {\bibfnamefont {A.}~\bibnamefont {Vinattieri}}, \bibinfo {author} {\bibfnamefont {J.}~\bibnamefont {Shah}}, \bibinfo {author} {\bibfnamefont {T.}~\bibnamefont {Damen}}, \bibinfo {author} {\bibfnamefont {D.}~\bibnamefont {Kim}}, \bibinfo {author} {\bibfnamefont {L.}~\bibnamefont {Pfeiffer}}, \bibinfo {author} {\bibfnamefont {M.}~\bibnamefont {Maialle}},\ and\ \bibinfo {author} {\bibfnamefont {L.}~\bibnamefont {Sham}},\ }\bibfield  {title} {\bibinfo {title} {Exciton dynamics in gaas quantum wells under resonant excitation},\ }\href {https://doi.org/10.1103/PhysRevB.50.10868} {\bibfield  {journal} {\bibinfo  {journal} {Phys. Rev. B}\ }\textbf {\bibinfo {volume} {50}},\ \bibinfo {pages} {10868} (\bibinfo {year} {1994})}\BibitemShut {NoStop}%
\bibitem [{\citenamefont {Eccleston}\ \emph {et~al.}(1991)\citenamefont {Eccleston}, \citenamefont {Strobel}, \citenamefont {R\"uhle}, \citenamefont {Kuhl}, \citenamefont {Feuerbacher},\ and\ \citenamefont {Ploog}}]{Eccleston1991_PRB}%
  \BibitemOpen
  \bibfield  {author} {\bibinfo {author} {\bibfnamefont {R.}~\bibnamefont {Eccleston}}, \bibinfo {author} {\bibfnamefont {R.}~\bibnamefont {Strobel}}, \bibinfo {author} {\bibfnamefont {W.}~\bibnamefont {R\"uhle}}, \bibinfo {author} {\bibfnamefont {J.}~\bibnamefont {Kuhl}}, \bibinfo {author} {\bibfnamefont {B.}~\bibnamefont {Feuerbacher}},\ and\ \bibinfo {author} {\bibfnamefont {K.}~\bibnamefont {Ploog}},\ }\bibfield  {title} {\bibinfo {title} {Exciton dynamics in a gaas quantum well},\ }\href {https://doi.org/10.1103/PhysRevB.44.1395} {\bibfield  {journal} {\bibinfo  {journal} {Phys. Rev. B}\ }\textbf {\bibinfo {volume} {44}},\ \bibinfo {pages} {1395} (\bibinfo {year} {1991})}\BibitemShut {NoStop}%
\bibitem [{\citenamefont {Hoyer}\ \emph {et~al.}(2005)\citenamefont {Hoyer}, \citenamefont {Ell}, \citenamefont {Kira}, \citenamefont {Koch}, \citenamefont {Chatterjee}, \citenamefont {Mosor}, \citenamefont {Khitrova}, \citenamefont {Gibbs},\ and\ \citenamefont {Stolz}}]{Hoyer2005_PRB}%
  \BibitemOpen
  \bibfield  {author} {\bibinfo {author} {\bibfnamefont {W.}~\bibnamefont {Hoyer}}, \bibinfo {author} {\bibfnamefont {C.}~\bibnamefont {Ell}}, \bibinfo {author} {\bibfnamefont {M.}~\bibnamefont {Kira}}, \bibinfo {author} {\bibfnamefont {S.}~\bibnamefont {Koch}}, \bibinfo {author} {\bibfnamefont {S.}~\bibnamefont {Chatterjee}}, \bibinfo {author} {\bibfnamefont {S.}~\bibnamefont {Mosor}}, \bibinfo {author} {\bibfnamefont {G.}~\bibnamefont {Khitrova}}, \bibinfo {author} {\bibfnamefont {H.}~\bibnamefont {Gibbs}},\ and\ \bibinfo {author} {\bibfnamefont {H.}~\bibnamefont {Stolz}},\ }\bibfield  {title} {\bibinfo {title} {Many-body dynamics and exciton formation studied by time-resolved photoluminescence},\ }\href {https://doi.org/10.1103/PhysRevB.72.075324} {\bibfield  {journal} {\bibinfo  {journal} {Phys. Rev. B}\ }\textbf {\bibinfo {volume} {72}},\ \bibinfo {pages} {075324} (\bibinfo {year} {2005})}\BibitemShut {NoStop}%
\bibitem [{\citenamefont {Cadiz}\ \emph {et~al.}(2018)\citenamefont {Cadiz}, \citenamefont {Robert}, \citenamefont {Courtade}, \citenamefont {Manca}, \citenamefont {Martinelli}, \citenamefont {Taniguchi}, \citenamefont {Watanabe}, \citenamefont {Amand}, \citenamefont {Rowe}, \citenamefont {Paget}, \citenamefont {Urbaszek},\ and\ \citenamefont {Marie}}]{Cadiz2018_APL}%
  \BibitemOpen
  \bibfield  {author} {\bibinfo {author} {\bibfnamefont {F.}~\bibnamefont {Cadiz}}, \bibinfo {author} {\bibfnamefont {C.}~\bibnamefont {Robert}}, \bibinfo {author} {\bibfnamefont {E.}~\bibnamefont {Courtade}}, \bibinfo {author} {\bibfnamefont {M.}~\bibnamefont {Manca}}, \bibinfo {author} {\bibfnamefont {L.}~\bibnamefont {Martinelli}}, \bibinfo {author} {\bibfnamefont {T.}~\bibnamefont {Taniguchi}}, \bibinfo {author} {\bibfnamefont {K.}~\bibnamefont {Watanabe}}, \bibinfo {author} {\bibfnamefont {T.}~\bibnamefont {Amand}}, \bibinfo {author} {\bibfnamefont {A.}~\bibnamefont {Rowe}}, \bibinfo {author} {\bibfnamefont {D.}~\bibnamefont {Paget}}, \bibinfo {author} {\bibfnamefont {B.}~\bibnamefont {Urbaszek}},\ and\ \bibinfo {author} {\bibfnamefont {X.}~\bibnamefont {Marie}},\ }\bibfield  {title} {\bibinfo {title} {Exciton diffusion in $\mathrm{WSe_2}$ monolayers embedded in a van der waals heterostructure},\ }\href {https://doi.org/10.1063/1.5026478} {\bibfield  {journal} {\bibinfo  {journal} {Appl. Phys. Lett.}\
  }\textbf {\bibinfo {volume} {112}},\ \bibinfo {pages} {152106} (\bibinfo {year} {2018})}\BibitemShut {NoStop}%
\bibitem [{\citenamefont {Snoke}(2011)}]{Snoke2011_AnnPhys}%
  \BibitemOpen
  \bibfield  {author} {\bibinfo {author} {\bibfnamefont {D.}~\bibnamefont {Snoke}},\ }\bibfield  {title} {\bibinfo {title} {The quantum boltzmann equation in semiconductor physics},\ }\href {https://doi.org/https://doi.org/10.1002/andp.201000102} {\bibfield  {journal} {\bibinfo  {journal} {Ann. Phys.}\ }\textbf {\bibinfo {volume} {523}},\ \bibinfo {pages} {87} (\bibinfo {year} {2011})}\BibitemShut {NoStop}%
\bibitem [{\citenamefont {Savenko}\ \emph {et~al.}(2011)\citenamefont {Savenko}, \citenamefont {Magnusson},\ and\ \citenamefont {Shelykh}}]{Savenko2011_PRB}%
  \BibitemOpen
  \bibfield  {author} {\bibinfo {author} {\bibfnamefont {I.}~\bibnamefont {Savenko}}, \bibinfo {author} {\bibfnamefont {E.}~\bibnamefont {Magnusson}},\ and\ \bibinfo {author} {\bibfnamefont {I.}~\bibnamefont {Shelykh}},\ }\bibfield  {title} {\bibinfo {title} {Density-matrix approach for an interacting polariton system},\ }\href {https://doi.org/10.1103/PhysRevB.83.165316} {\bibfield  {journal} {\bibinfo  {journal} {Phys. Rev. B}\ }\textbf {\bibinfo {volume} {83}},\ \bibinfo {pages} {165316} (\bibinfo {year} {2011})}\BibitemShut {NoStop}%
\bibitem [{\citenamefont {Snoke}\ \emph {et~al.}(1991)\citenamefont {Snoke}, \citenamefont {Braun},\ and\ \citenamefont {Cardona}}]{Snoke1991_PRB}%
  \BibitemOpen
  \bibfield  {author} {\bibinfo {author} {\bibfnamefont {D.}~\bibnamefont {Snoke}}, \bibinfo {author} {\bibfnamefont {D.}~\bibnamefont {Braun}},\ and\ \bibinfo {author} {\bibfnamefont {M.}~\bibnamefont {Cardona}},\ }\bibfield  {title} {\bibinfo {title} {Carrier thermalization in $\mathrm{Cu_2O}$: Phonon emission by excitons},\ }\href {https://doi.org/10.1103/PhysRevB.44.2991} {\bibfield  {journal} {\bibinfo  {journal} {Phys. Rev. B}\ }\textbf {\bibinfo {volume} {44}},\ \bibinfo {pages} {2991} (\bibinfo {year} {1991})}\BibitemShut {NoStop}%
\bibitem [{\citenamefont {Snoke}\ \emph {et~al.}(1992)\citenamefont {Snoke}, \citenamefont {Shields},\ and\ \citenamefont {Cardona}}]{Snoke1992_PRB}%
  \BibitemOpen
  \bibfield  {author} {\bibinfo {author} {\bibfnamefont {D.}~\bibnamefont {Snoke}}, \bibinfo {author} {\bibfnamefont {A.}~\bibnamefont {Shields}},\ and\ \bibinfo {author} {\bibfnamefont {M.}~\bibnamefont {Cardona}},\ }\bibfield  {title} {\bibinfo {title} {Phonon-absorption recombination luminescence of room-temperature excitons in $\mathrm{Cu_2O}$},\ }\href {https://doi.org/10.1103/PhysRevB.45.11693} {\bibfield  {journal} {\bibinfo  {journal} {Phys. Rev. B}\ }\textbf {\bibinfo {volume} {45}},\ \bibinfo {pages} {11693} (\bibinfo {year} {1992})}\BibitemShut {NoStop}%
\bibitem [{\citenamefont {O'Hara}\ and\ \citenamefont {Wolfe}(2000)}]{OHara2000_PRB}%
  \BibitemOpen
  \bibfield  {author} {\bibinfo {author} {\bibfnamefont {K.}~\bibnamefont {O'Hara}}\ and\ \bibinfo {author} {\bibfnamefont {J.}~\bibnamefont {Wolfe}},\ }\bibfield  {title} {\bibinfo {title} {Relaxation kinetics of excitons in cuprous oxide},\ }\href {https://doi.org/10.1103/PhysRevB.62.12909} {\bibfield  {journal} {\bibinfo  {journal} {Phys. Rev. B}\ }\textbf {\bibinfo {volume} {62}},\ \bibinfo {pages} {12909} (\bibinfo {year} {2000})}\BibitemShut {NoStop}%
\bibitem [{\citenamefont {Ivanov}\ \emph {et~al.}(1997)\citenamefont {Ivanov}, \citenamefont {Ell},\ and\ \citenamefont {Haug}}]{Ivanov1997_PRE}%
  \BibitemOpen
  \bibfield  {author} {\bibinfo {author} {\bibfnamefont {A.}~\bibnamefont {Ivanov}}, \bibinfo {author} {\bibfnamefont {C.}~\bibnamefont {Ell}},\ and\ \bibinfo {author} {\bibfnamefont {H.}~\bibnamefont {Haug}},\ }\bibfield  {title} {\bibinfo {title} {Phonon-assisted boltzmann kinetics of a bose gas: Generic solution for $t \leq t_c$},\ }\href {https://doi.org/10.1103/PhysRevE.55.6363} {\bibfield  {journal} {\bibinfo  {journal} {Phys. Rev. E}\ }\textbf {\bibinfo {volume} {55}},\ \bibinfo {pages} {6363} (\bibinfo {year} {1997})}\BibitemShut {NoStop}%
\bibitem [{\citenamefont {Potma}\ and\ \citenamefont {Wiersma}(1998)}]{Potma1998_JChemPhys}%
  \BibitemOpen
  \bibfield  {author} {\bibinfo {author} {\bibfnamefont {E.}~\bibnamefont {Potma}}\ and\ \bibinfo {author} {\bibfnamefont {D.}~\bibnamefont {Wiersma}},\ }\bibfield  {title} {\bibinfo {title} {Exciton superradiance in aggregates: The effect of disorder, higher order exciton-phonon coupling and dimensionality},\ }\href {https://doi.org/10.1063/1.475898} {\bibfield  {journal} {\bibinfo  {journal} {J. Chem. Phys.}\ }\textbf {\bibinfo {volume} {108}},\ \bibinfo {pages} {4894} (\bibinfo {year} {1998})}\BibitemShut {NoStop}%
\bibitem [{\citenamefont {Selig}\ \emph {et~al.}(2018)\citenamefont {Selig}, \citenamefont {Bergh\"auser}, \citenamefont {Richter}, \citenamefont {Bratschitsch}, \citenamefont {Knorr},\ and\ \citenamefont {Malic}}]{Selig2018_2DMat}%
  \BibitemOpen
  \bibfield  {author} {\bibinfo {author} {\bibfnamefont {M.}~\bibnamefont {Selig}}, \bibinfo {author} {\bibfnamefont {G.}~\bibnamefont {Bergh\"auser}}, \bibinfo {author} {\bibfnamefont {M.}~\bibnamefont {Richter}}, \bibinfo {author} {\bibfnamefont {R.}~\bibnamefont {Bratschitsch}}, \bibinfo {author} {\bibfnamefont {A.}~\bibnamefont {Knorr}},\ and\ \bibinfo {author} {\bibfnamefont {E.}~\bibnamefont {Malic}},\ }\bibfield  {title} {\bibinfo {title} {Dark and bright exciton formation, thermalization, and photoluminescence in monolayer transition metal dichalcogenides},\ }\href {https://doi.org/10.1088/2053-1583/aabea3} {\bibfield  {journal} {\bibinfo  {journal} {2D Mater.}\ }\textbf {\bibinfo {volume} {5}},\ \bibinfo {pages} {035017} (\bibinfo {year} {2018})}\BibitemShut {NoStop}%
\bibitem [{\citenamefont {Chen}\ \emph {et~al.}(2022)\citenamefont {Chen}, \citenamefont {Sangalli},\ and\ \citenamefont {Bernardi}}]{Chen2022_PhysRevRes}%
  \BibitemOpen
  \bibfield  {author} {\bibinfo {author} {\bibfnamefont {H.}~\bibnamefont {Chen}}, \bibinfo {author} {\bibfnamefont {D.}~\bibnamefont {Sangalli}},\ and\ \bibinfo {author} {\bibfnamefont {M.}~\bibnamefont {Bernardi}},\ }\bibfield  {title} {\bibinfo {title} {First-principles ultrafast exciton dynamics and time-domain spectroscopies: Dark-exciton mediated valley depolarization in monolayer ${\mathrm{wse}}_{2}$},\ }\href {https://doi.org/10.1103/PhysRevResearch.4.043203} {\bibfield  {journal} {\bibinfo  {journal} {Phys. Rev. Research}\ }\textbf {\bibinfo {volume} {4}},\ \bibinfo {pages} {043203} (\bibinfo {year} {2022})}\BibitemShut {NoStop}%
\bibitem [{\citenamefont {Piermarocchi}\ \emph {et~al.}(1997)\citenamefont {Piermarocchi}, \citenamefont {Tassone}, \citenamefont {Savona}, \citenamefont {Quattropani},\ and\ \citenamefont {Schwendimann}}]{Piermarocchi1997_PRB}%
  \BibitemOpen
  \bibfield  {author} {\bibinfo {author} {\bibfnamefont {C.}~\bibnamefont {Piermarocchi}}, \bibinfo {author} {\bibfnamefont {F.}~\bibnamefont {Tassone}}, \bibinfo {author} {\bibfnamefont {V.}~\bibnamefont {Savona}}, \bibinfo {author} {\bibfnamefont {A.}~\bibnamefont {Quattropani}},\ and\ \bibinfo {author} {\bibfnamefont {P.}~\bibnamefont {Schwendimann}},\ }\bibfield  {title} {\bibinfo {title} {Exciton formation rates in {G}a{A}s/{A}l$_{x}${G}a$_{1-x}${A}s quantum wells},\ }\href {https://doi.org/10.1103/PhysRevB.55.1333} {\bibfield  {journal} {\bibinfo  {journal} {Phys. Rev. B}\ }\textbf {\bibinfo {volume} {55}},\ \bibinfo {pages} {1333} (\bibinfo {year} {1997})}\BibitemShut {NoStop}%
\bibitem [{\citenamefont {Zhang}\ \emph {et~al.}(1997)\citenamefont {Zhang}, \citenamefont {Huang},\ and\ \citenamefont {Zhou}}]{Zhang1997_JPCM}%
  \BibitemOpen
  \bibfield  {author} {\bibinfo {author} {\bibfnamefont {M.}~\bibnamefont {Zhang}}, \bibinfo {author} {\bibfnamefont {Q.}~\bibnamefont {Huang}},\ and\ \bibinfo {author} {\bibfnamefont {J.}~\bibnamefont {Zhou}},\ }\bibfield  {title} {\bibinfo {title} {Calculations of the time taken for excitons to form in gaas quantum wells},\ }\href {https://doi.org/10.1088/0953-8984/9/46/016} {\bibfield  {journal} {\bibinfo  {journal} {J. Phys. Condens. Matter.}\ }\textbf {\bibinfo {volume} {9}},\ \bibinfo {pages} {10185} (\bibinfo {year} {1997})}\BibitemShut {NoStop}%
\bibitem [{\citenamefont {Piermarocchi}\ \emph {et~al.}(1996)\citenamefont {Piermarocchi}, \citenamefont {Tassone}, \citenamefont {Savona}, \citenamefont {Quattropani},\ and\ \citenamefont {Schwendimann}}]{Piermarocchi1996_PRB}%
  \BibitemOpen
  \bibfield  {author} {\bibinfo {author} {\bibfnamefont {C.}~\bibnamefont {Piermarocchi}}, \bibinfo {author} {\bibfnamefont {F.}~\bibnamefont {Tassone}}, \bibinfo {author} {\bibfnamefont {V.}~\bibnamefont {Savona}}, \bibinfo {author} {\bibfnamefont {A.}~\bibnamefont {Quattropani}},\ and\ \bibinfo {author} {\bibfnamefont {P.}~\bibnamefont {Schwendimann}},\ }\bibfield  {title} {\bibinfo {title} {Nonequilibrium dynamics of free quantum-well excitons in time-resolved photoluminescence},\ }\href {https://doi.org/10.1103/PhysRevB.53.15834} {\bibfield  {journal} {\bibinfo  {journal} {Phys. Rev. B}\ }\textbf {\bibinfo {volume} {53}},\ \bibinfo {pages} {15834} (\bibinfo {year} {1996})}\BibitemShut {NoStop}%
\bibitem [{\citenamefont {Selbmann}\ \emph {et~al.}(1996)\citenamefont {Selbmann}, \citenamefont {Gulia}, \citenamefont {Rossi}, \citenamefont {Molinari},\ and\ \citenamefont {Lugli}}]{Selbmann1996_PRB}%
  \BibitemOpen
  \bibfield  {author} {\bibinfo {author} {\bibfnamefont {P.}~\bibnamefont {Selbmann}}, \bibinfo {author} {\bibfnamefont {M.}~\bibnamefont {Gulia}}, \bibinfo {author} {\bibfnamefont {F.}~\bibnamefont {Rossi}}, \bibinfo {author} {\bibfnamefont {E.}~\bibnamefont {Molinari}},\ and\ \bibinfo {author} {\bibfnamefont {P.}~\bibnamefont {Lugli}},\ }\bibfield  {title} {\bibinfo {title} {Coupled free-carrier and exciton relaxation in optically excited semiconductors},\ }\href {https://doi.org/10.1103/PhysRevB.54.4660} {\bibfield  {journal} {\bibinfo  {journal} {Phys. Rev. B}\ }\textbf {\bibinfo {volume} {54}},\ \bibinfo {pages} {4660} (\bibinfo {year} {1996})}\BibitemShut {NoStop}%
\bibitem [{\citenamefont {Golub}\ \emph {et~al.}(1996)\citenamefont {Golub}, \citenamefont {Scherbakov},\ and\ \citenamefont {Akimov}}]{Golub1996_JPCM}%
  \BibitemOpen
  \bibfield  {author} {\bibinfo {author} {\bibfnamefont {L.}~\bibnamefont {Golub}}, \bibinfo {author} {\bibfnamefont {A.}~\bibnamefont {Scherbakov}},\ and\ \bibinfo {author} {\bibfnamefont {A.}~\bibnamefont {Akimov}},\ }\bibfield  {title} {\bibinfo {title} {Energy distributions of 2d excitons in the presence of nonequilibrium phonons},\ }\href {https://doi.org/10.1088/0953-8984/8/13/008} {\bibfield  {journal} {\bibinfo  {journal} {J. Phys. Condens. Matter.}\ }\textbf {\bibinfo {volume} {8}},\ \bibinfo {pages} {2163} (\bibinfo {year} {1996})}\BibitemShut {NoStop}%
\bibitem [{\citenamefont {Basu}\ and\ \citenamefont {Ray}(1992)}]{Basu1992_PRB}%
  \BibitemOpen
  \bibfield  {author} {\bibinfo {author} {\bibfnamefont {P.}~\bibnamefont {Basu}}\ and\ \bibinfo {author} {\bibfnamefont {P.}~\bibnamefont {Ray}},\ }\bibfield  {title} {\bibinfo {title} {Energy relaxation of hot two-dimensional excitons in a gaas quantum well by exciton-phonon interaction},\ }\href {https://doi.org/10.1103/PhysRevB.45.1907} {\bibfield  {journal} {\bibinfo  {journal} {Phys. Rev. B}\ }\textbf {\bibinfo {volume} {45}},\ \bibinfo {pages} {1907} (\bibinfo {year} {1992})}\BibitemShut {NoStop}%
\bibitem [{\citenamefont {Lee}\ \emph {et~al.}(1986)\citenamefont {Lee}, \citenamefont {Koteles},\ and\ \citenamefont {Vassell}}]{Lee1986_PRB}%
  \BibitemOpen
  \bibfield  {author} {\bibinfo {author} {\bibfnamefont {J.}~\bibnamefont {Lee}}, \bibinfo {author} {\bibfnamefont {E.}~\bibnamefont {Koteles}},\ and\ \bibinfo {author} {\bibfnamefont {M.}~\bibnamefont {Vassell}},\ }\bibfield  {title} {\bibinfo {title} {Luminescence linewidths of excitons in gaas quantum wells below 150 k},\ }\href {https://doi.org/10.1103/PhysRevB.33.5512} {\bibfield  {journal} {\bibinfo  {journal} {Phys. Rev. B}\ }\textbf {\bibinfo {volume} {33}},\ \bibinfo {pages} {5512} (\bibinfo {year} {1986})}\BibitemShut {NoStop}%
\bibitem [{\citenamefont {Combescot}\ and\ \citenamefont {Betbeder-Matibet}(2002)}]{Combescot2002_EurohysLett}%
  \BibitemOpen
  \bibfield  {author} {\bibinfo {author} {\bibfnamefont {M.}~\bibnamefont {Combescot}}\ and\ \bibinfo {author} {\bibfnamefont {O.}~\bibnamefont {Betbeder-Matibet}},\ }\bibfield  {title} {\bibinfo {title} {The effective bosonic hamiltonian for excitons reconsidered},\ }\href {https://doi.org/10.1209/epl/i2002-00609-3} {\bibfield  {journal} {\bibinfo  {journal} {Europhys. Lett.}\ }\textbf {\bibinfo {volume} {58}},\ \bibinfo {pages} {87} (\bibinfo {year} {2002})}\BibitemShut {NoStop}%
\bibitem [{\citenamefont {Combescot}\ \emph {et~al.}(2008{\natexlab{a}})\citenamefont {Combescot}, \citenamefont {Betbeder-Matibet},\ and\ \citenamefont {Dubin}}]{Combescot2008_PhysRep}%
  \BibitemOpen
  \bibfield  {author} {\bibinfo {author} {\bibfnamefont {M.}~\bibnamefont {Combescot}}, \bibinfo {author} {\bibfnamefont {O.}~\bibnamefont {Betbeder-Matibet}},\ and\ \bibinfo {author} {\bibfnamefont {F.}~\bibnamefont {Dubin}},\ }\bibfield  {title} {\bibinfo {title} {The many-body physics of composite bosons},\ }\href {https://doi.org/https://doi.org/10.1016/j.physrep.2007.11.003} {\bibfield  {journal} {\bibinfo  {journal} {Physics Reports}\ }\textbf {\bibinfo {volume} {463}},\ \bibinfo {pages} {215} (\bibinfo {year} {2008}{\natexlab{a}})}\BibitemShut {NoStop}%
\bibitem [{\citenamefont {Combescot}\ \emph {et~al.}(2007)\citenamefont {Combescot}, \citenamefont {Betbeder-Matibet},\ and\ \citenamefont {Combescot}}]{Combescot2007_PRB}%
  \BibitemOpen
  \bibfield  {author} {\bibinfo {author} {\bibfnamefont {M.}~\bibnamefont {Combescot}}, \bibinfo {author} {\bibfnamefont {O.}~\bibnamefont {Betbeder-Matibet}},\ and\ \bibinfo {author} {\bibfnamefont {R.}~\bibnamefont {Combescot}},\ }\bibfield  {title} {\bibinfo {title} {Exciton-exciton scattering: Composite boson versus elementary boson},\ }\href {https://doi.org/10.1103/PhysRevB.75.174305} {\bibfield  {journal} {\bibinfo  {journal} {Phys. Rev. B}\ }\textbf {\bibinfo {volume} {75}},\ \bibinfo {pages} {174305} (\bibinfo {year} {2007})}\BibitemShut {NoStop}%
\bibitem [{\citenamefont {Thilagam}(2015)}]{Thilagam2015_APA}%
  \BibitemOpen
  \bibfield  {author} {\bibinfo {author} {\bibfnamefont {A.}~\bibnamefont {Thilagam}},\ }\bibfield  {title} {\bibinfo {title} {Influence of the pauli exclusion principle on scattering properties of cobosons},\ }\href {https://doi.org/https://doi.org/10.1016/j.physb.2014.10.021} {\bibfield  {journal} {\bibinfo  {journal} {Phys. B: Condens. Matter.}\ }\textbf {\bibinfo {volume} {457}},\ \bibinfo {pages} {232} (\bibinfo {year} {2015})}\BibitemShut {NoStop}%
\bibitem [{\citenamefont {Katzer}\ \emph {et~al.}(2023)\citenamefont {Katzer}, \citenamefont {Selig}, \citenamefont {Sigl}, \citenamefont {Troue}, \citenamefont {Figueiredo}, \citenamefont {Kiemle}, \citenamefont {Sigger}, \citenamefont {Wurstbauer}, \citenamefont {Holleitner},\ and\ \citenamefont {Knorr}}]{Katzer2023_arxiv}%
  \BibitemOpen
  \bibfield  {author} {\bibinfo {author} {\bibfnamefont {M.}~\bibnamefont {Katzer}}, \bibinfo {author} {\bibfnamefont {M.}~\bibnamefont {Selig}}, \bibinfo {author} {\bibfnamefont {L.}~\bibnamefont {Sigl}}, \bibinfo {author} {\bibfnamefont {M.}~\bibnamefont {Troue}}, \bibinfo {author} {\bibfnamefont {J.}~\bibnamefont {Figueiredo}}, \bibinfo {author} {\bibfnamefont {J.}~\bibnamefont {Kiemle}}, \bibinfo {author} {\bibfnamefont {F.}~\bibnamefont {Sigger}}, \bibinfo {author} {\bibfnamefont {U.}~\bibnamefont {Wurstbauer}}, \bibinfo {author} {\bibfnamefont {A.}~\bibnamefont {Holleitner}},\ and\ \bibinfo {author} {\bibfnamefont {A.}~\bibnamefont {Knorr}},\ }\href@noop {} {\bibinfo {title} {Exciton-phonon-scattering: A competition between bosonic and fermionic nature of bound electron-hole pairs}} (\bibinfo {year} {2023}),\ \Eprint {https://arxiv.org/abs/2303.11787} {arXiv:2303.11787 [cond-mat.mes-hall]} \BibitemShut {NoStop}%
\bibitem [{\citenamefont {Carmichael}(2007)}]{Carmichael2007}%
  \BibitemOpen
  \bibfield  {author} {\bibinfo {author} {\bibfnamefont {H.}~\bibnamefont {Carmichael}},\ }\href@noop {} {\emph {\bibinfo {title} {Quantum Optics 1: Master Equations and Fokker-Planck Equations}}}\ (\bibinfo  {publisher} {Springer},\ \bibinfo {address} {New York},\ \bibinfo {year} {2007})\BibitemShut {NoStop}%
\bibitem [{\citenamefont {Kavokin}\ and\ \citenamefont {Malpuech}(2003)}]{Kavokin2003}%
  \BibitemOpen
  \bibfield  {author} {\bibinfo {author} {\bibfnamefont {A.}~\bibnamefont {Kavokin}}\ and\ \bibinfo {author} {\bibfnamefont {G.}~\bibnamefont {Malpuech}},\ }\href@noop {} {\emph {\bibinfo {title} {Cavity polaritons}}}\ (\bibinfo  {publisher} {Elsevier Academic Press},\ \bibinfo {address} {Amsterdam},\ \bibinfo {year} {2003})\BibitemShut {NoStop}%
\bibitem [{\citenamefont {Combescot}\ \emph {et~al.}(2008{\natexlab{b}})\citenamefont {Combescot}, \citenamefont {Betbeder-Matibet},\ and\ \citenamefont {Dubin}}]{COMBESCOT2008215}%
  \BibitemOpen
  \bibfield  {author} {\bibinfo {author} {\bibfnamefont {M.}~\bibnamefont {Combescot}}, \bibinfo {author} {\bibfnamefont {O.}~\bibnamefont {Betbeder-Matibet}},\ and\ \bibinfo {author} {\bibfnamefont {F.}~\bibnamefont {Dubin}},\ }\bibfield  {title} {\bibinfo {title} {The many-body physics of composite bosons},\ }\href {https://doi.org/https://doi.org/10.1016/j.physrep.2007.11.003} {\bibfield  {journal} {\bibinfo  {journal} {Phys. Rep.}\ }\textbf {\bibinfo {volume} {463}},\ \bibinfo {pages} {215} (\bibinfo {year} {2008}{\natexlab{b}})}\BibitemShut {NoStop}%
\bibitem [{\citenamefont {Agranovich}\ and\ \citenamefont {Toshich}(1968)}]{Agranovich1968}%
  \BibitemOpen
  \bibfield  {author} {\bibinfo {author} {\bibfnamefont {V.~M.}\ \bibnamefont {Agranovich}}\ and\ \bibinfo {author} {\bibfnamefont {B.~S.}\ \bibnamefont {Toshich}},\ }\bibfield  {title} {\bibinfo {title} {Collective properties of frenkel excitons},\ }\href@noop {} {\bibfield  {journal} {\bibinfo  {journal} {Soviet Physics JETP}\ }\textbf {\bibinfo {volume} {26}},\ \bibinfo {pages} {104} (\bibinfo {year} {1968})},\ \bibinfo {note} {translated from Zh. Eksp. Teor. Fiz. 53, 149--162 (1967)}\BibitemShut {NoStop}%
\bibitem [{\citenamefont {Kira}\ and\ \citenamefont {Koch}(2006)}]{KIRA2006155}%
  \BibitemOpen
  \bibfield  {author} {\bibinfo {author} {\bibfnamefont {M.}~\bibnamefont {Kira}}\ and\ \bibinfo {author} {\bibfnamefont {S.}~\bibnamefont {Koch}},\ }\bibfield  {title} {\bibinfo {title} {Many-body correlations and excitonic effects in semiconductor spectroscopy},\ }\href {https://doi.org/https://doi.org/10.1016/j.pquantelec.2006.12.002} {\bibfield  {journal} {\bibinfo  {journal} {Prog. Quantum Electron.}\ }\textbf {\bibinfo {volume} {30}},\ \bibinfo {pages} {155} (\bibinfo {year} {2006})}\BibitemShut {NoStop}%
\bibitem [{\citenamefont {Madelung}(2004)}]{Madelung2004}%
  \BibitemOpen
  \bibfield  {author} {\bibinfo {author} {\bibfnamefont {O.}~\bibnamefont {Madelung}},\ }\href {https://doi.org/10.1007/978-3-642-18865-7} {\emph {\bibinfo {title} {Semiconductors: Data Handbook}}}\ (\bibinfo  {publisher} {Springer-Verlag, Berlin Heidelberg},\ \bibinfo {year} {2004})\BibitemShut {NoStop}%
\end{thebibliography}%

\end{document}